\documentclass[graybox,natbib,nosecnum]{svmult}
\bibpunct{(}{)}{;}{a}{}{,} 

\pdfoutput=1   

\usepackage{amssymb}        
\usepackage{mathptmx}       
\usepackage{helvet}         
\usepackage{courier}        
\usepackage{type1cm}        

\usepackage{makeidx}         
\usepackage{graphicx}        
\usepackage{multicol}        
\usepackage[bottom]{footmisc}
\usepackage[normalem]{ulem}	
\usepackage{hyperref}  

\usepackage{soul}   

\newcommand*\aap{A\&A}

\newcommand*\aj{AJ}

\newcommand*\apj{ApJ}
\newcommand*\apjl{ApJ}

\newcommand*\araa{ARA\&A}

\newcommand*\icarus{Icarus}

\newcommand*\mnras{MNRAS}

\newcommand*\nar{New A Rev}
\newcommand*\nat{Nature}

\newcommand*\pasp{PASP}

\newcommand*\psj{Planet Sci J}

\newcommand*\ssr{Space Sci Rev}

\newcommand{\hbindex}[1]{{#1}}  

\makeindex             

\begin{document}

\title*{Highlights from Exoplanet Observations by the James Webb Space Telescope}
\author{N\'estor Espinoza  and Marshall Perrin}
\institute{N\'estor Espinoza \at Space Telescope Science Institute, 3700 San Martin Drive, Baltimore, MD 21218, USA \at Department of Physics and Astronomy, Johns Hopkins University, Baltimore, MD 21218, USA \email{nespinoza@stsci.edu}
\and Marshall Perrin \at Space Telescope Science Institute, 3700 San Martin Drive, Baltimore, MD 21218, USA \email{mperrin@stsci.edu}}
%
%
\maketitle

\abstract{The \textit{James Webb Space Telescope} (\textit{JWST}) has started a revolution in exoplanetary science. From studying in exquisite detail the chemical inventories and physical processes in gas giant exoplanets, the structure and chemical diversity of the enigmatic sub-Neptune population to even providing constraints on the atmospheric make-up of rocky exoplanets, the observatory is enabling cutting-edge science that is touching virtually every sub-area in the field. In this review Chapter, we showcase key highlights from exoplanet science being conducted with this state-of-the-art space observatory, which we believe is representative of the transformational science it is producing. One of the key takeaways from these pioneering \textit{JWST} observations is how they are starting to reshape not only how we think, study and interpret exoplanet observations --- but how they are also reshaping our intuition about our very own Solar System planets.}

\section{Introduction}

The \textit{\hbindex{James Webb Space Telescope}} \citep[\textit{\hbindex{JWST}};][]{jwst2023} is a revolutionary mission for exoplanet science. With its four science instruments, it enables photometry and spectroscopy of exoplanets from 0.6 to 27.9 $\mu$m with unprecedented sensitivity and stability. This allows the observatory to perform a wide variety of exoplanet studies, from directly detecting light from young, recently formed planetary mass objects to exploring the atmospheres of small, potentially habitable rocky worlds --- a versatility that has been instrumental to its popularity and success within the field \cite[see][for a review of the observatory's capabilities and instrumentation for exoplanet science]{handbook-chapter-jwst}.


The birth of stars and planetary systems, and the characterization of planetary systems were two of the four major science themes identified during the mission's development. It should not comes as a surprise, thus, that \textit{JWST} exoplanet science observations comprise, as of Cycle 4, around $30\%$ of all science observations, making this field one of the main drivers of the observatory's scientific operations. At the time of writing --- 2.5 years after the start of scientific operations --- \textit{JWST} data have led to around 150 refereed publications which span a wide array of science cases. The observatory is still early in its expected mission lifetime, and we expect this number to keep growing rapidly in the upcoming years.

In this review chapter, we aim to highlight results from the mission that are representative of the transformational science that the observatory is yielding. We focus primarily on observations to discover and characterize exoplanets, touching only lightly on the observing programs for the closely-related topics of planetary system formation and circumstellar disks. For the purposes of this review and after careful consideration, we have decided to \textit{loosely} follow the International Astronomical Union (IAU) definition of an exoplanet \citep{desetangs:2022} and consider bound objects with masses lower than about 12 Jupiter-masses an exoplanet. We do not explicitly apply the IAU's strict definition on the mass ratio between the orbiting object and the central object being smaller than 1/25, however. In our usage, thus, TWA 27 B \citep{chauvin:2004} is considered an exoplanet, but, e.g., WISE 0855–0714 \citep{luhman:2014} is not. We do note this is done with the sole purpose of containing the material in this chapter within the limits of a few tens of pages --- we strongly believe, as prior observations have suggested and \textit{JWST} itself is confirming, all those objects clearly form a continuum that needs to be studied as such. 

\subsection{Overview of JWST Exoplanet Observations}

The vast majority of exoplanet observations with the observatory are carried out via two main techniques: \textbf{high-contrast} and \textbf{transiting exoplanet time-series} imaging/spectroscopy. In both of these techniques, the main thrust in these early years has been the \textit{detailed characterization of already-known exoplanets} discovered by other facilities. Recently there have begun to be results in \textit{discovery of new exoplanets} only detectable by JWST thus far. 

\textbf{\hbindex{High-contrast imaging/spectroscopy}} is typically used when exoplanets are, or are expected to be, resolved as point sources by the observatory. These techniques are used to study not only planetary mass objects --- here referred to as ``direct" imaging/spectroscopy of exoplanets --- but also protoplanetary, debris and even circumplanetary disks. The term ``high-contrast" stems from the need to maximize the dynamic range of the images and/or spectra enough to detect the faint glow of exoplanets embedded in the bright glare dominated by the flux of the stellar host. This typically requires flux ratio detections of order $\sim 10^{-5}$ at relatively small separations ($\lesssim 1$ arcseconds). To achieve this, detailed point-spread function (PSF) subtraction/modeling techniques are typically employed \citep[see, e.g., ][]{kammerer:2022, carter:2023, ruffio:2024}. In addition, \textit{JWST} is also equipped with starlight suppression technology in some instruments that can be used to block a portion of the starlight. In particular, both the MIRI \citep{miri-coron} and NIRCam \citep{nircam-coron} instruments onboard \textit{JWST} possess coronagraphs, while the NIRISS instrument has a 7-hole non-redundant mask (NRM) that allows it to perform Aperture Masking Interferometry \citep[AMI;][]{niriss-ami} --- a first in space-based instrumentation. However, JWST's two imaging spectrographs, the NIRSpec IFU and MIRI MRS, cannot be used with the coronagaphs and therefore must rely entirely on data processing to achieve the necessary high contrast for imaging spectroscopy of exoplanets.  For direct imaging, typical science targets are Jovian planets at separations of 10s of au from their host star, and often with a preference for younger planetary systems in which the contrast ratios are more favorable. JWST's primary advantages over prior facilities for direct imaging are (1) much deeper sensitivity, particularly at wavelengths $>3 \mu$m, (2) wide wavelength coverage spanning the infrared wavelengths at which planetary atmospheres are brightest, and (3) superb optical stability enabling high-precision PSF subtractions. Compared to current state-of-the-art ground facilities on 8-10 m telescopes, however, it can be more difficult for JWST to access planets at the smallest angular separations.

\textbf{\hbindex{Transiting exoplanet} time-series imaging/spectroscopy}, on the other hand, is typically used when exoplanets themselves are not spatially resolved by the observatory but are known to transit their host stars. Spectroscopy which allows atmospheric characterization of these exoplanets can be achieved by targeting primary transits (when the exoplanet transits in front of the star), secondary eclipses (when the exoplanet passes behind the star) or even phase changes of the exoplanet as it orbits around its host star, which manifests as flux variations in time --- a ``phase-curve" \citep[see, e.g., ][for a detailed overview of these techniques]{kreidberg:2018, parmentier:2018, alonso:2018}.  These require extreme relative flux stability in time in order to detect the wavelength-dependent flux variations of these events at the 10-100 part-per-million (ppm) level. This is achieved by \textit{JWST} on the one hand by the extreme stability of the observatory as a whole \citep{lajoie:2023}, but also thanks to specific decisions on the instrument design that minimize the possibility of time-dependent systematic effects. These range from large slits or slitless designs to avoid possible slit losses due to pointing jitter \citep[see, e.g., the cases of MIRI and NIRSpec;][]{kendrew:2015, birkmann:2022, wright:2023} to special optical elements and readout electronics that allow for precise, high-cadence time-series observations \citep[see, e.g., the case for NIRCam and NIRISS;][]{schlawin:2017, rieke:2023, albert:2023}. The typical science targets for time series observations span from Jovian planets to rocky Earth-mass planets, the majority of which orbit at small separations with typical orbital periods from hours to days. \textit{JWST}'s primary advantages over prior facilities for time series observations are (1) its superb observatory and instrumental stability and precision, (2) the wide wavelength coverage spanning key spectral regions for exoplanetary science, and (3) a very capable instrument suite with options for both broad spectral coverage and relatively high spectral resolution depending on science needs. 



\begin{figure}
\includegraphics[scale=0.45]{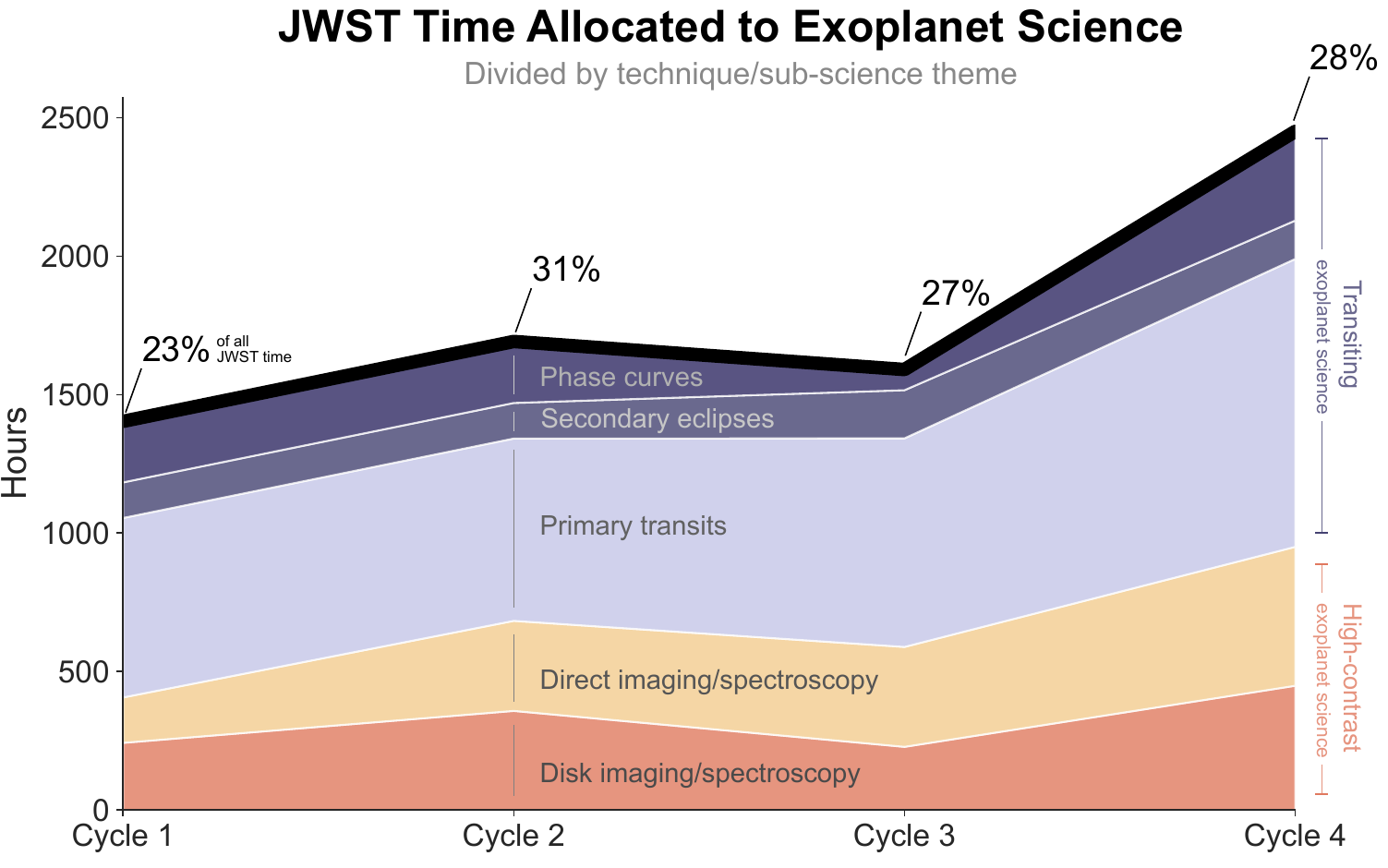}
\caption{\textbf{JWST GO time allocation for exoplanet science on different Cycles.} Representation of the total General Observer (GO) allocated time for exoplanet science on \textit{JWST} Cycle 1 (released on March, 2021, i.e., pre-launch), Cycle 2 (released on May 2023), Cycle 3 (released on February, 2024) and Cycle 4 (released on March, 2025). The percentage of all telescope time spent on exoplanet science for a given Cycle is indicated on top of this illustration (e.g., Cycle 1 comprised 23\% of time on the telescope devoted to exoplanet science). The purple shading corresponds to the share allocated to transiting exoplanet science while the orange shading corresponds to high-contrast exoplanet science performed with \textit{JWST}. }
\label{fig:1}       
\end{figure}

\subsection{JWST Exoplanet Targets and Observing Programs}

\begin{figure}
\includegraphics[scale=0.45]{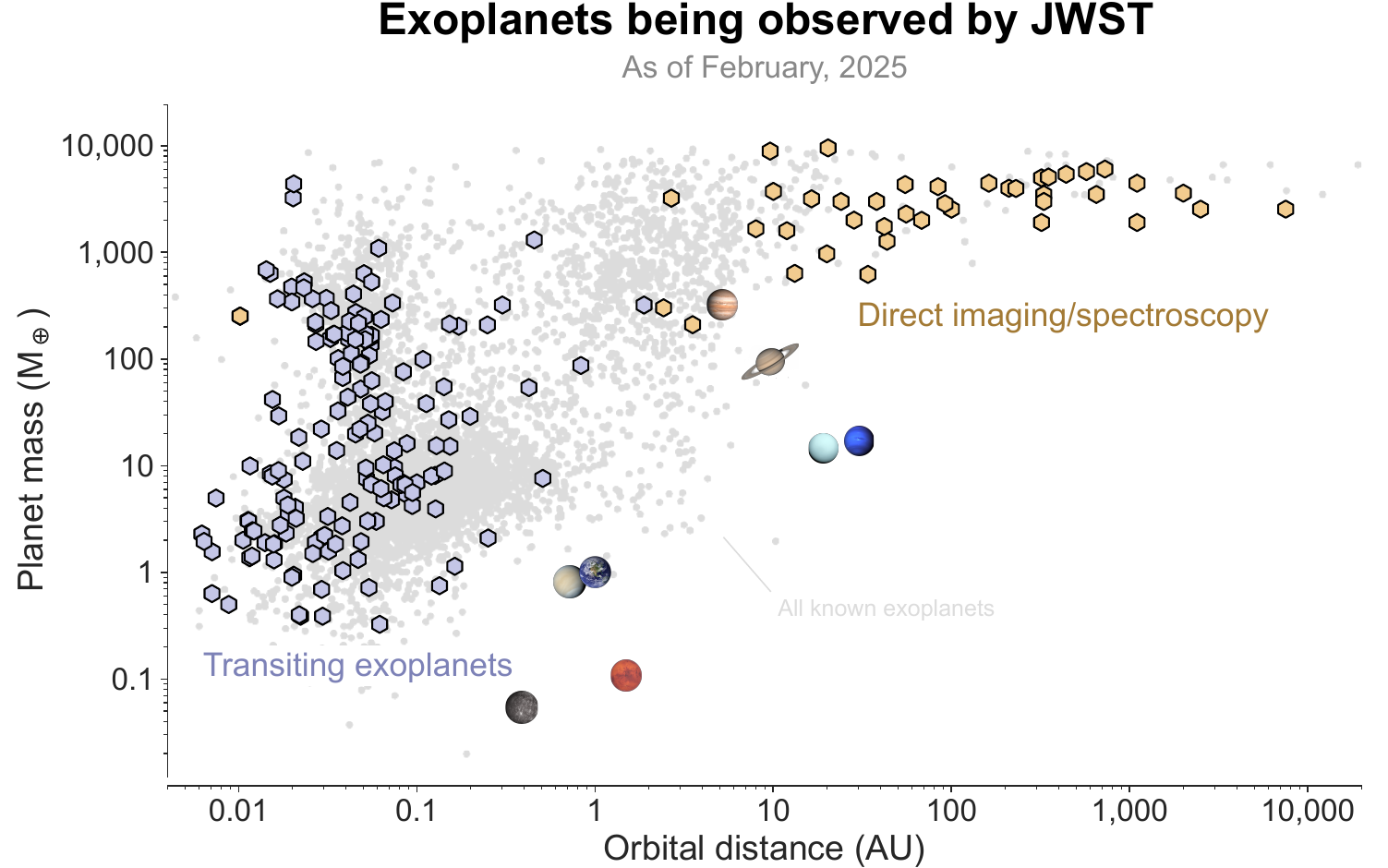}
\caption{\textbf{Exoplanets being observed by JWST as of February 2025.} Orbital distance in Astronomical Units (AU) versus planetary mass (in Earth's units) of known exoplanets being observed by JWST through February 2025 (hexagons). This includes Guaranteed Time Observations (GTO), Director's Discretionary Time (DDT), Early Release Observations (ERO), Early Release Science (ERS) and General Observer (GO) observations from Cycles 1, 2 and 3. Yellow hexagons represent JWST direct imaging/spectroscopy exoplanet observations, while purple hexagon represent transiting exoplanet observations with JWST. Grey points represent all known confirmed exoplanets. Images of Solar System planets are embedded for reference.}
\label{fig:2}       
\end{figure}

To illustrate how the exoplanet community is using \textit{JWST} time between those two techniques, we present in Figure \ref{fig:1} the time allocation of \hbindex{\textit{JWST} exoplanet science} proposed through the General Observer's (GO) program for its first four Cycles. On average, around $\sim 1,000$ hours per Cycle ($\sim 1,500$ hours in Cycle 4) are devoted to transiting exoplanet science, with the bulk of the time being used to target primary transit events, while around $\sim 500$ hours per Cycle ($\sim 950$ hours in Cycle 4) are devoted to direct imaging/spectroscopy of exoplanets and disks. The majority of these programs focus on in-depth studies of already known exoplanets. Figure \ref{fig:2} places these \textit{JWST} targeted worlds in context of all known exoplanets and objects of our own Solar System, shown in the classic orbital distance versus planetary mass diagram. At the time of writing, this represents \textit{JWST} observations of over 200 exoplanets orbiting over 160 stars in total. 

The distribution of \hbindex{exoplanets being observed by \textit{JWST}} in Figure \ref{fig:2} helps explain the popularity of transiting exoplanet science in terms of \textit{JWST} time allocation: there is a larger pool of targets to apply these techniques to. These, in turn, span a wide range of planetary masses from terrestrial to gas giants embedded in a variety of stellar environments, from close-in gas giants orbiting very hot A-stars \citep[e.g., observations of WASP-178 b;][]{lothringer-prop} to the numerous --- hundreds of hours of --- \textit{JWST} observations of the terrestrial exoplanets orbiting the ultra-cool dwarf TRAPPIST-1 \citep{guillon:2024}. The vast majority of these observations aim to perform detailed atmospheric characterization of these distant worlds. This exploration has produced benchmark spectra of highly-irradiated, Jupiter-sized objects revealing never-before-seen molecules in their atmospheres such as CO$_2$ and SO$_2$ \citep[see, e.g.,][]{co2:2023, tsai:2023, carter:2023}, are allowing the detailed atmospheric characterization of exoplanets with no analogue in the Solar System such as the enigmatic sub-Neptune population \citep[see, e.g.,][]{madhusudhan:2023, benneke:2024, holmberg:2024} and are even enabling studies aiming to unveil possible atmospheres from terrestrial worlds in the habitable zone of their stars \citep[see, e.g.,][]{neat:2017, lewis-prop:2017, lim-prop:2021, allen:2024}. 

While the pool of accessible targets is currently not as large for high-contrast exoplanet observations, and until very recently only included young ($\lesssim 1$ Gyr) exoplanetary systems, that technique accesses a unique parameter space in terms of the more widely-separated and massive known exoplanets, allowing to study a wide variety of phenomena. Studies range from measuring the emergent flux of currently forming young exoplanetary systems, which holds key clues about their formation \citep[see, e.g.,][]{christiaens:2024, blakely:2024}, to unveiling the detailed chemical inventories in the atmospheres of objects at the boundary of what actually \textit{is considered} an exoplanet \citep[see, e.g.,][]{miles:2023}. In addition, a number of \textit{JWST} high-contrast imaging exoplanet programs are aimed at discovering \textit{new} exoplanetary systems. This is enabled both by the wide range of infrared wavelengths \textit{JWST} gives access to and to the high contrast ratios at small separations attainable by the stability and instrumentation onboard the observatory. This has recently enabled the first imaging detections of more ``mature" giant exoplanet candidates. These recent results range from candidate exoplanets orbiting white dwarfs \citep[see, e.g.,][]{mullally:2024, limbach:2024}, key to understand the future of planetary systems such as our own Solar System, to exoplanets around main-sequence stars, which hold the promise to help put our Solar System giant planets in context with other planetary systems elsewhere \citep{matthews:2024}. Ongoing \textit{JWST} programs are expected to enable the detection of exoplanets even below the mass of Saturn, which would open up a new parameter space in mass for directly imaged exoplanets --- key to complete the puzzle to explain substructures on disks in young systems \citep[see, e.g.,][]{huges:2018, andrews:2020, vdm:2023}, as well as to draw parallels to our very own Solar System planets \citep[see, e.g.,][]{beichman-prop:2021, marino-prop:2021, lagrange-prop:2023, carter-prop:2023, carter-prop:2024}.

While the above overview of \hbindex{\textit{JWST} exoplanet observations} summarize the majority of observations being made on exoplanetary systems, it omits a number of observing programs performing innovative exoplanet science that does not fit those descriptions. For example, the program led by \cite{zhang-prop:2024} is aiming to perform the very first study on the atmospheric make-up of a planetary mass object orbiting a pulsar --- PSR-J2322-2650~b \citep{spiewak:2018} --- by directly detecting light from the exoplanet, given the pulsar is expected to be virtually invisible at infrared wavelengths. The programs led by \cite{cassese-prop:2024} and \cite{pass-prop:2024} are, on the other hand, aiming to constrain the presence of satellites --- ``exomoons" \citep[see, e.g.,][for a review]{exomoons:2024} --- orbiting a Jupiter-analogue and a terrestrial exoplanet in the habitable zone of its star, respectively, taking advantage of the exquisite photometric precision enabled by \textit{JWST}, which could reveal transits of these satellites in front of the host star and/or impacts on the observed transits of the host planets themselves. Another interesting example is that of the programs of \cite{vanCapelleveen-prop:2023} and \cite{kenworthy-prop:2024}, which are aiming to study what could be the remnants of a planetary collision --- a \textit{synestia} --- which \cite{kenworthy:2023} reveals might be orbiting the young star ASASSN-21qj. As can be observed, the extreme on-sky precision of the \textit{JWST} observatory is enabling scientific ideas that might not have been thought of when it was being designed, but which are undoubtedly expanding the very limit of our knowledge of exoplanetary systems.

\subsection{Structure of this Chapter}

Having presented an overview of the kind of observations \textit{JWST} is performing on exoplanetary systems, we now move to presenting highlights from these observations in the published literature. While biased, we believe these represent significant advances of our understanding of exoplanets enabled by the \textit{JWST} mission --- at least up to the time of writing of this document (March 2025). We elected to divide these highlights by planetary size and mass. We begin with highlights on the observations of gas giants, which covers observations of planets from the masses of Neptune all the way to super-Jupiters, and includes observations performed with both the high-contrast and transiting exoplanet techniques. Then, we present highlights on the observations of exoplanets smaller than Neptune all the way down to terrestrial exoplanetary science which, as can be seen in Figure \ref{fig:2}, are mainly being studied by \textit{JWST} via transiting exoplanets. First, we focus our presentation of highlights on sub-Neptune exoplanets, which have no analogue in our Solar System and which, at least for periods less than 100 days, are the most numerous population in terms of their occurrence rate \citep[see, e.g.,][and references therein]{fp:2018}. Finally, we discuss highlights on the observation of rocky exoplanets by \textit{JWST}, which will cover observations of objects going from super-Earths all the way down to Earth-sized exoplanets. We end this Chapter with a Conclusions section which summarizes these highlights.

\section{Highlights of \textit{JWST} observations of gas giant exoplanets}

\hbindex{Gas giant exoplanets} (i.e., exoplanets with masses $\gtrsim 10 M_\oplus$) have been the main focus of exoplanet in-depth exploration for at least the first few decades of exoplanet exploration, kickstarted by the discovery of the highly irradiated Jupiter-mass exoplanet 51 Pegasi~b \citep{mq:1995}. The reasons for their popularity are numerous. From an observational perspective, gas giant exoplanets provide the best signal-to-noise ratios for both the high-contrast and transiting exoplanet techniques due to their large sizes and the wide range of orbital distances these exoplanets have been found to orbit. This, in turn, allows their study in a wide variety of stellar environments and irradiation levels. From a theoretical perspective, the detection of these objects via high-constrast imaging provides key constraints on their occurrence at large separations \citep{bowler:2018}, opening in the way new targets to characterize in detail. This characterization provides in turn a unique window to a variety of chemical inventories, which can be used to understand currently ongoing chemical and dynamical processes in their atmospheres \citep[see, e.g.,][and references therein]{moses:2011, moses:2016, gao:2021, tsai:2023}, as well as to unveil unique planet formation and evolution imprints in their envelopes \citep[see, e.g.,][and references therein]{oberg:2011, mordasini:2016, venturini:2016, thorngren:2016, espinoza:2017, madhusudhan:2017}. In this Section, we highlight how \textit{JWST} is exploring these and other scientific questions with its exquisite instrumentation. 



\subsection{\textit{JWST}'s unveiling of gas giant chemical inventories}

Prior to \textit{JWST}, the exploration of the chemical inventories of \hbindex{gas giant exoplanet atmospheres} was pioneered by ground-based and \textit{Hubble Space Telescope} (\textit{HST}) observations of exoplanets \citep[see, e.g.,][for a review]{deming-seager:2017}. For direct imaging, this exploration has been a fruitful one in particular from the ground, with spectra at near-infrared wavelengths ($1-2.5$ $\mu$m) producing a variety of detections of H$_2$O, CO and even hints of CH$_4$ on massive, young super-Jupiters \citep[see, e.g.,][and references therein]{biller:2018, chauvin:2024}. For transiting exoplanets, while high-resolution ground-based observations have been very productive at detecting atomic features on the brightest of targets attainable from current observatories \citep[see, e.g.,][for a recent example]{prinoth:2025}, constraints on molecular abundances have been mostly elusive for molecules other than H$_2$O, routinely detected using \textit{HST}'s Wide Field Camera 3 (WFC3) near-infrared capabilities \citep[1-1.7 $\mu$m; see, e.g.,][]{sing:2016}. For both techniques, constraints on molecular abundances beyond about 2 $\mu$m resorted in large part on photometric points which provided clues but not definitive constraints on important absorbers at longer wavelengths such as CH$_4$, a key tracer of chemical disequilibrium processes and CO$_2$, fundamental to constrain atmospheric metallicity on (exo)planets \citep{lodders:2002}. \textit{JWST} observations, thus, were highly anticipated to complete the picture of chemical inventories in gas giant exoplanets. 

\subsubsection{Benchmark spectra from Early Release Science (ERS) Programs}
\begin{figure}
\includegraphics[scale=0.45]{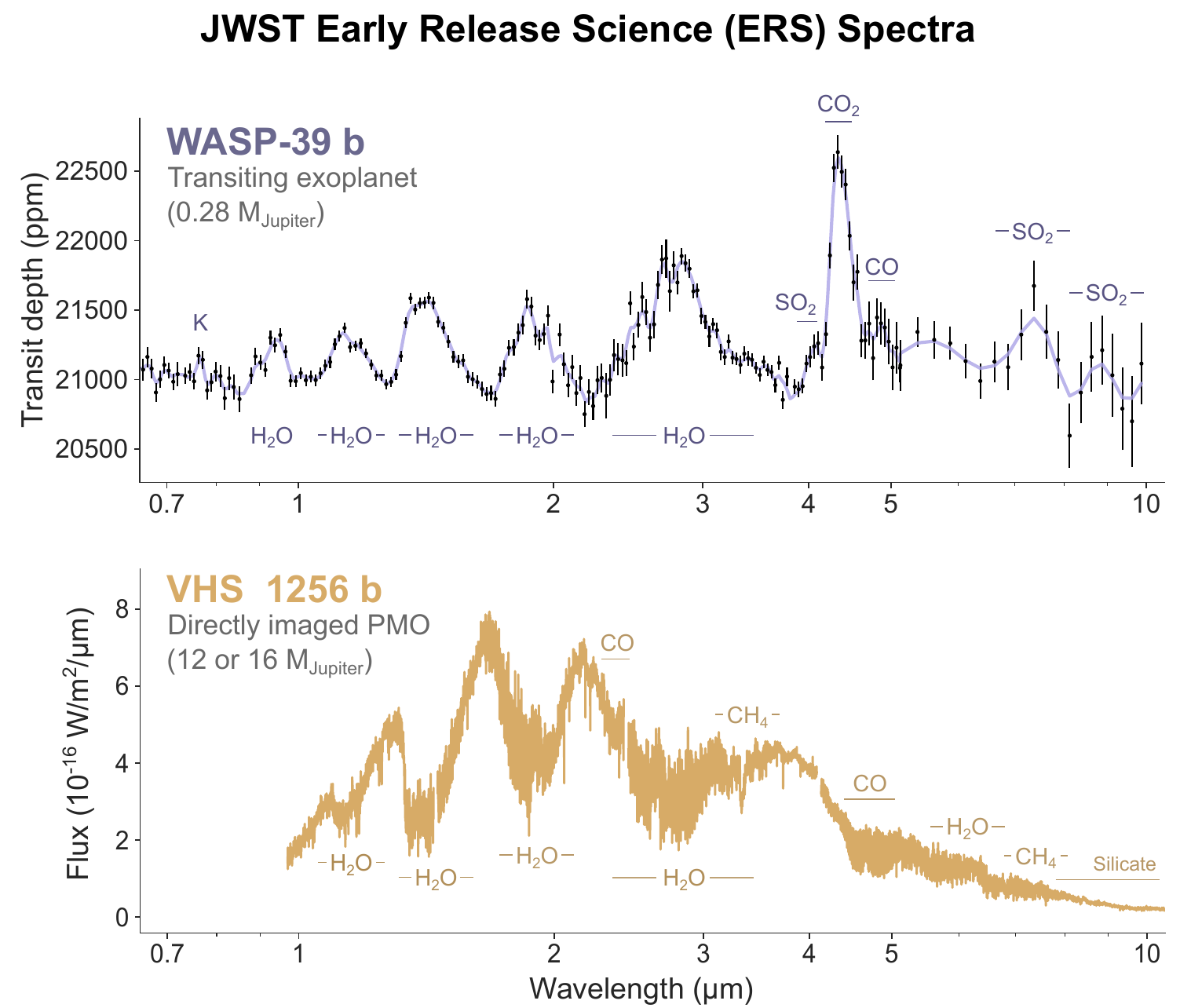}
\caption{\textbf{\textit{JWST}-quality spectra of transiting and directly-imaged exoplanets.} Transmission spectrum of the transiting exoplanet WASP-39~b (top, black points --- purple line is a gaussian filter to aid the eye) and emergent spectrum of the directly imaged planetary mass object (PMO) VHS 1256~b (bottom, yellow) obtained through the instrumentation onboard \textit{JWST} shown here up to 10 $\mu$m, for comparison. Both objects have similar temperatures (around 1200 K), but vastly different masses (labeled under each object's name). Note how the objects share the location of certain features (e.g., H$_2$O), but not others (e.g., lack of CH$_4$ but presence of photochemical products like SO$_2$ in WASP-39~b). Feature labeled as Silicate in VHS 1256~b corresponds to a silicate cloud feature in its atmosphere. Data for WASP-39~b from \cite{carter-may:2023} up to 5 $\mu$m and \cite{powell:2024} from 5-10 $\mu$m. Data for VHS 1256~b from \cite{miles:2023}. }
\label{fig:3}       
\end{figure}

The very first set of detailed spectra of exoplanets with \textit{JWST} more than delivered on the observatory expectations during its first few months of scientific operations. One such set were produced as part of two STScI Director’s Discretionary \hbindex{Early Release Science} (DD ERS) programs: the program focused on transiting exoplanet science of \cite{ers:2017} and the program focused on high-contrast imaging and spectroscopy of exoplanets led by \cite{ers-hc:2017}. We showcase the stunning spectra produced by those teams in Figure \ref{fig:3}: a panchromatic spectrum of the highly irradiated ($T_{\textrm{eq}}\approx 1200$ K), Saturn-mass ($0.28$ M$_\textrm{Jupiter}$) transiting exoplanet WASP-39~b, here produced by combining observations from  NIRISS/SOSS \citep{feinstein:2023}, NIRCam/Grism \citep{ahrer:2023} and NIRSpec/G395H \citep{alderson:2023} which forms the spectrum from $0.65-5$ $\mu$m compiled and uniformly analyzed by \cite{carter-may:2023}, and the panchromatic spectrum of the young and hot ($T_{\textrm{eff}} \sim 1200$ K) directly imaged planetary mass object (12-16 M$_\textrm{Jupiter}$) VHS 1256~b presented in \citep{miles:2023}, which is in turn also a compilation of spectra over the same instruments plus MIRI. We have also included in Figure \ref{fig:3} the 5-10 $\mu$m spectrum obtained by \cite{powell:2024} for WASP-39~b using the MIRI/LRS instrument.

As detailed in \cite{miles:2023}, the spectra of VHS 1256~b is nothing short of revolutionary. While muted with respect to chemical equilibrium model expectations, the prominent absorption feature of CH$_4$ at 3.3 $\mu$m already showcases the impact of \hbindex{vertical mixing} and clouds in the atmosphere of VHS 1256~b. The possibility of observing this molecule together with CO features --- which also trace \hbindex{disequilibrium processes} --- in the same spectrum is a first, and a possibility enabled by the wide range of wavelengths the observatory allows to study. The fact that in addition a strong feature of silicate clouds is observed in the spectrum at wavelengths $\gtrsim 8$ $\mu$m further marks this spectrum as one that will be studied for years to come as a benchmark spectrum of a directly imaged planetary mass object. Indeed, studies performing in-depth explorations of bulk planetary parameters enabled to be extracted from this spectrum are already beginning to showcase the difficulty of fitting the plethora of features present in it \cite[see, e.g.,][]{petrus:2024,lueber:2024}, while others are showcasing how this exquisite data even enables the detection of CO isotopologues \citep{gandhi:2023}.

WASP-39~b's transmission spectrum is also a game changer. As detailed in the studies of the transiting exoplanet ERS team \citep{co2:2023, rustamkulov:2023, alderson:2023}, the observed CO$_2$ feature at 4.3 $\mu$m marks the very first time this molecule has been unambiguously detected in a planet beyond our Solar System, a detection enabled by \textit{JWST}'s exquisite spectrophotometric precision and wavelength coverage. In turn, this feature provides a strong constraint on the metallicity of WASP-39~b \citep[about $\times$10 solar;][]{rustamkulov:2023}. The detection of several H$_2$O features which don't exactly follow predictions from equilibrium chemistry calculations, as described in \cite{feinstein:2023} provide in addition strong evidence for inhomogeneous cloud coverage on the exoplanet, a prediction from General Circulation Models \citep[GCMs;][]{carone:2023}. One of the most striking outcomes from this spectroscopic exploration of WASP-39~b, however, was an initially unexpected feature at about 4 $\mu$m which was later identified by \cite{tsai:2023} as arising from \hbindex{sulfur dioxide} (SO$_2$); a feature confirmed by the follow-up observations of \cite{powell:2024} with MIRI/LRS (5-10 $\mu$m; annotated in Figure \ref{fig:3}). This once again marks a first: this is the very first time we are able to constrain \hbindex{photochemical processes} in planets outside our Solar System, which were predicted to be present in highly irradiated gas giant exoplanets such as WASP-39~b \citep{zahnle:2009, polman:2023}. Just as in the case of VHS 1256~b, this benchmark spectra is giving rise to a wide range of studies, from detailed constraints on WASP-39~b's bulk properties to detailed explorations of its chemical makeup \citep[see, e.g.,][]{niraula:2023, constantinou:2023, khorshid:2024, lueber-w39:2024, al:2024, esparza-borges:2023, grant-co:2023}.

\subsubsection{Highlights of atomic and molecular constraints on gas giants by \textit{JWST}}

\begin{figure}
\includegraphics[scale=0.45]{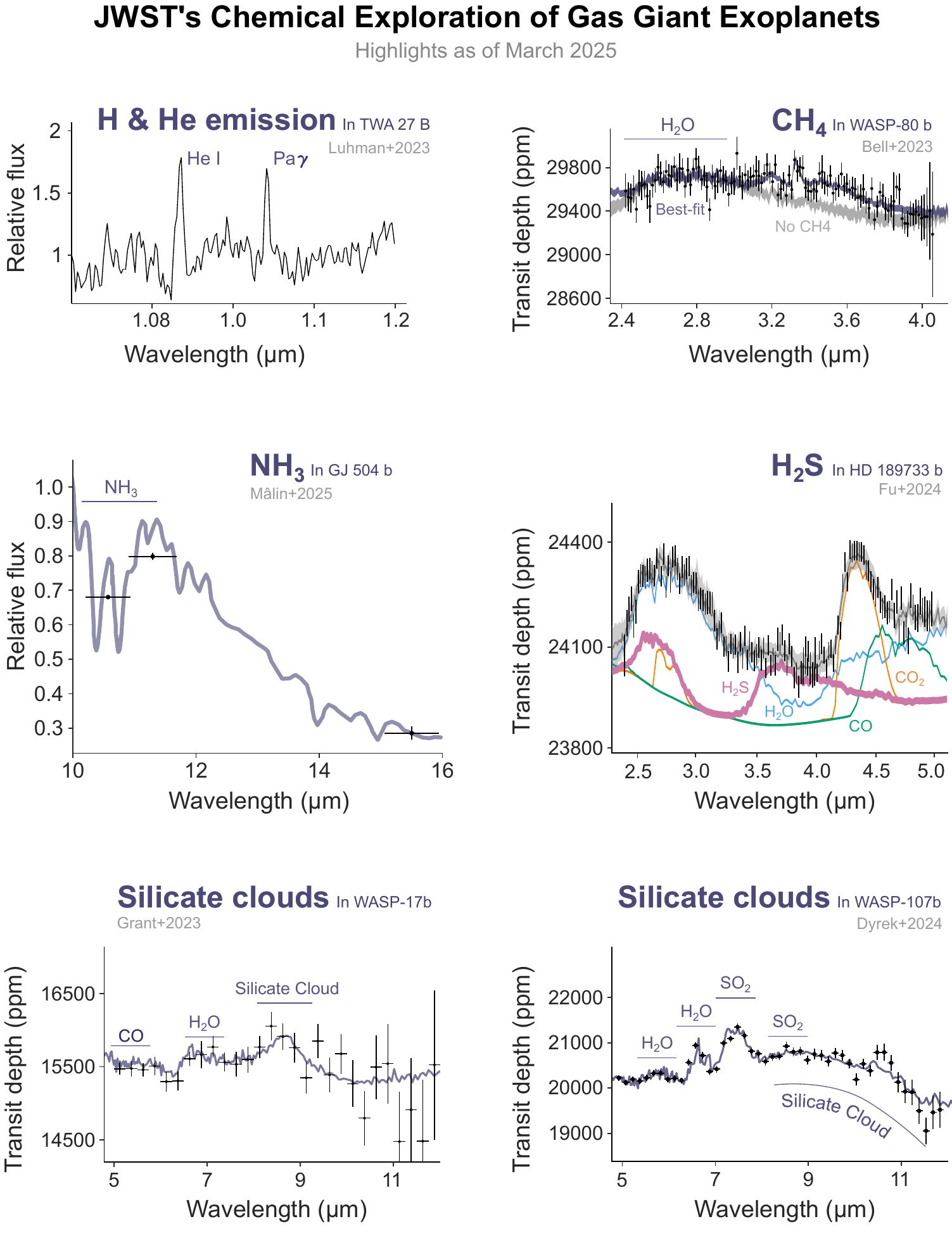}
\caption{\textbf{Some highlights on \textit{JWST}'s chemical exploration of gas giant exoplanets.} Transmission and emergent flux for transiting and directly imaged exoplanets, respectively, showing spectroscopic features that are paving the way in the exploration of \textit{JWST} gas giant exoplanet atmospheric exploration. (\textit{Top left}) Helium and hydrogen emission features observed in the emergent spectrum of the 5 M$_{\textnormal{Jupiter}}$ planetary mass object TWA 27 B via NIRSpec high-resolution spectroscopy; adapted from \cite{luhman:2023}. (\textit{Top right}) The first detection of CH$_4$ in low-resolution transmission spectroscopy for a Saturn-mass transiting exoplanet via NIRCam/Grism observations; adapted from \cite{bell:2023}. (\textit{Middle left}) The first unambiguous detection of NH$_3$ in a planetary mass object via MIRI coronagraphy on the directly imaged 5 M$_{\textnormal{Jupiter}}$ GJ 504~b; adapted from \cite{malin:2025}, under a Creative Commons Attribution (CC BY 4.0) license. (\textit{Middle right}) The first tight constraint on H$_2$S for the Jupiter-mass transiting exoplanet HD 189733~b via NIRCam/Grism observations; adapted from \cite{fu:2024}.  (\textit{Bottom}) The first spectroscopic signatures of clouds for Jupiter-to-Neptune mass, transiting gas giants: for the Hot Jupiter WASP-17~b and for WASP-107~b using MIRI/LRS observations; adapted from \cite{grant:2023} and \cite{dyrek:2024}. }
\label{fig:4}       
\end{figure}

The spectroscopic features presented and studied by the ERS teams acted as teasers of the state-of-the-art capabilities \textit{JWST} has for exoplanet atmospheric exploration. Features such as the ones showcased in those spectra are now routinely being detected on both directly imaged \citep[see, e.g., ][]{luhman:2023, manjavacas:2024, worthen:2024} and transiting \citep[see, e.g.,][and references therein]{fu:2025} exoplanets along with an even \textit{larger} variety of chemical compounds with an unprecedented level of detail. 

Among the highlights of this detailed chemical exploration of gas giant exoplanets are the ones defining key advances in our understanding of young, massive planetary mass objects. One example of this is the detection of the He I triplet, a set of Paschen lines (Pa$\alpha$, Pa$\beta$, Pa$\gamma$ and Pa$\delta$) and even hints of Brackett-series lines \textit{in emission} in the 5 Jupiter-mass, directly imaged planetary mass object TWA 27 B via \textit{JWST} NIRSpec observations \citep[see Figure \ref{fig:4}, top left panel;][]{luhman:2023, marleau:2024}. The detection of these clear signatures of accretion marks the very first time a wide number of H/He lines in emission are detected for such a low mass planetary mass object to date \citep[see, e.g.,][]{betti:2023}. This, in turn, is enabling detailed studies of the accretion process on TWA 27 B, from its geometry to its origins \citep{marleau:2024, aoyama:2024}. Another example is the detection of NH$_3$ in the 5 Jupiter-mass planetary mass object GJ 504~b via \textit{JWST} MIRI coronagraphic multi-band photometry. This discovery opens the door to high-precision nitrogen chemistry constraints on planetary mass objects, which is key for precise constraints on gravities — and thus masses, in objects like GJ 504~b  \cite[see Figure \ref{fig:4}, middle left panel;][]{malin:2025}. 

The chemical exploration for lower mass gas giant exoplanets --- being mainly done through transiting exoplanets --- is pushing new exciting directions as well. One key set of results in the last few years has been the detection of methane both in the 0.5 Jupiter-mass, warm (800 K) transiting exoplanet WASP-80~b, observed with \textit{JWST} NIRCam both in emission (i.e., in the exoplanet's dayside) and transmission \cite[i.e., in the exoplanet's terminator --- see Figure \ref{fig:4}, top right panel;][]{bell:2023} and its detection in the 2 Neptune-mass, warm (700 K) transiting exoplanet WASP-107~b by two independent teams via transmission spectroscopy using two different instruments: \textit{JWST} NIRCam \citep{welbanks:2024} and NIRSpec \citep{sing:2016}. These set of studies mark the first time relatively high precision constraints on the abundance of CH$_4$ have been made for Neptune-to-Jupiter mass exoplanets, kickstarting the exploration of the long-awaited and predicted constraints this molecule can provide, including insights on \hbindex{vertical mixing} and internal temperatures \citep[see, e.g.,][and references therein]{fortney:2020}. This is exactly what was enabled for WASP-107~b, on which the 3 order of magnitude depletion in CH$_4$ with respect to chemical equilibrium predictions allowed the teams to reveal strong vertical mixing, with \hbindex{eddy diffusion} coefficients $K_{ZZ} \gtrsim 10^{8}$ cm$^2$ s$^{-1}$ \citep{sing:2024, welbanks:2024} ---similar or higher than the largest values assumed in the most turbulent parts of the atmospheres of gas and ice giants in the Solar System \citep{teanby:2020}--- and a high internal temperature ($>345$ K) for the exoplanet --- about 3 times larger than the expected value given the planet's mass and age, likely due to \hbindex{eccentricity-driven tidal heating} \citep{welbanks:2024}. 

Another set of results that are important to highlight involves the first solid detection of H$_2$S in the transmission spectrum of the hot (1200 K) Jupiter-mass exoplanet HD 189733~b, performed with \textit{JWST} NIRCam and studied in detail by two independent teams \citep[Figure \ref{fig:4}, middle right panel;][]{fu:2024, zhang:2025}. While H$_2$S is in general predicted to be the dominant sulfur carrier in thermochemical equilibrium calculations for hot ($\sim 1000$ K), H/He-dominated atmospheres, its photochemically-triggered destruction can significantly lower its abundance at pressures relevant for transmission spectroscopy ($\lesssim 0.1$ mbar). This destruction, in turn, leads to the significant increase of SO$_2$ at those high-altitudes, sometimes even making the latter the dominant sulfur carrier: the very effect that explains its detection in WASP-39~b \citep{polman:2023, tsai:2023, crossfield:2023}. Theoretical studies show that while photochemically-produced SO$_2$ depends strongly on planetary bulk parameters such as temperature and C/O ratios, H$_2$S is relatively stable to those \citep[see, e.g.,][and references therein]{polman:2023}. H$_2$S, thus, although difficult to detect \citep{polman:2023}, could act as an anchor for the planet's inherent sulfur abundance --- a window studies such as the ones performed in HD 189733~b by \cite{fu:2024} and \cite{zhang:2025} have now opened.
 
The last highlight on the chemical exploration of gas giant exoplanets we discuss is the detection of spectroscopic signatures of clouds in the Jupiter-mass exoplanet HD 189733~b \citep{inglis:2024}, the $0.5$ Jupiter-mass hot (1700 K) Jupiter WASP-17~b \citep[Figure \ref{fig:4}, bottom left panel;][]{grant:2023}, and on the 2 Neptune-mass exoplanet WASP-107~b \citep[Figure \ref{fig:4}, bottom right panel;][]{dyrek:2024} via \textit{JWST} MIRI/LRS observations --- a first at those long wavelengths for such low mass gas giant exoplanets. From the very first observations of transiting exoplanet atmospheres, \hbindex{clouds} have been shown to damp spectroscopic features in transmission \citep{charbonneau:2002}, and to be overall ubiquitous on gas giant exoplanets \citep[see, e.g.,][and references therein]{sing:2016, wakef:2019, brande:2024}. However, the composition of these clouds has remained a mystery as predicted spectroscopic features of different cloud compositions are mostly located at infrared wavelengths ($>5\ \mu$m), beyond the detection capabilities of any observatory prior to \textit{JWST} \citep[see, e.g.,][and references therein]{wakeford:2015}. The detection of these long-awaited and predicted features, thus, has opened a new window into cloud physics thanks to \textit{JWST} --- with quartz clouds (i.e., SiO$_2$ condensates) being the most likely explanation so far for the observed spectroscopic features in all those works \citep{grant:2023,dyrek:2024,inglis:2024}. 

\subsection{Gas giant exoplanets in multiple dimensions}

\begin{figure}
\includegraphics[scale=0.45]{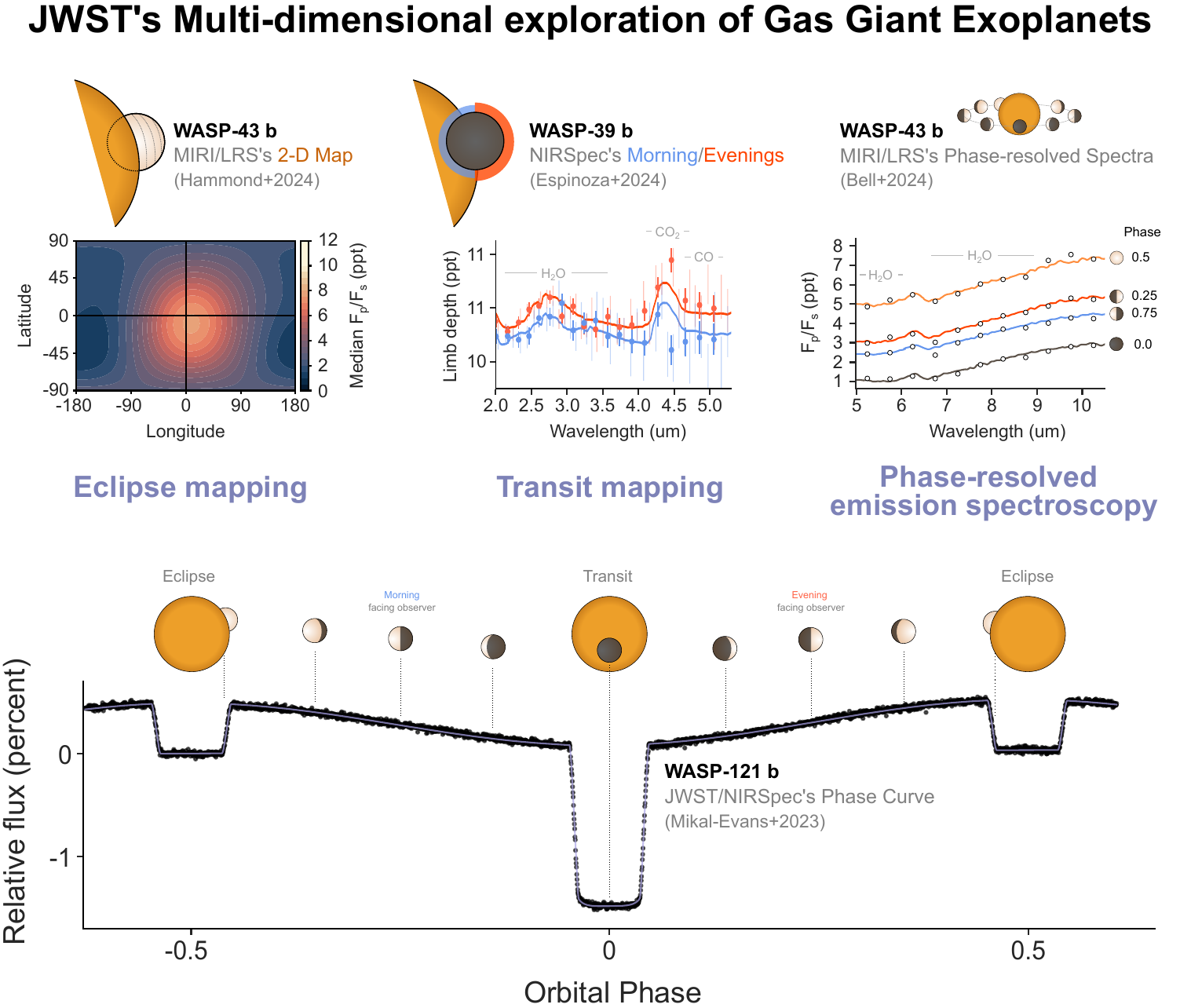}
\caption{\textbf{Some highlights on \textit{JWST}'s multi-dimensional exploration of gas giant exoplanets.} (\textit{Bottom}) Anotated, raw NIRSpec/G395H NRS2 phase-curve of WASP-121~b --- \textit{JWST}'s first phase curve event published in the refereed literature, showcasing the exquisite precision of the observatory to study these events \citep[adapted from][]{mikal-evans:2023}. To match the geometry of the transit mapping illustration (morning/evening), the illustration of phases throughout the phase-curve shown on top of it show the planet as it moved clockwise around its star (assumed to be the same direction of the rotation of the planet). (\textit{Top, left}) Flux map of WASP-43~b derived from fitting eclipse maps to its MIRI/LRS secondary eclipse lightcurve, which allows to obtain both longitudinal \textit{and} latitudinal information about the planet's flux distribution \citep[adapted from][]{hammond:2024}. 
 (\textit{Top, middle}) Morning/evening spectrum of WASP-39~b obtained by separating the terminator components using NIRSpec/PRISM wavelength-dependent transit events \citep[adapted from][under a Creative Commons Attribution (CC BY) license]{espinoza:2024}. (\textit{Top, right}) Phase-resolved emission spectroscopy of WASP-43~b, obtained by studying its MIRI/LRS phase curve --- best-fit retrievals to each phase are plotted on top of the datapoints \citep[adapted from][under a Creative Commons Attribution (CC BY) license]{bell:2024}}
\label{fig:5}       
\end{figure}

Exoplanets are ---at the very least--- 3-dimensional objects. This implies properties such as those unveiled by \textit{JWST} discussed in the previous sub-sections might actually vary across both longitude and latitude. Pioneering studies from both ground and space-based observatories have indeed explored constraints on those processes, with ample success in particular from space-based observatories at infrared wavelengths such as \textit{HST} and \textit{Spitzer} \citep[see, e.g.,][and references therein]{knutson:2007, majeau:2012, dewit:2012, stevenson:2014}. For gas giant exoplanets, these studies kickstarted a very productive decade of 3D constraints on these worlds, from important constraints on the day-to-night temperature contrasts of several tens of highly irradiated transiting giant exoplanets via phase-curves with 3.6 and 4.5$\mu$m \textit{Spitzer} photometry \citep[see, e.g.,][and references therein]{dang:2025} to studies of the wavelength-dependent emission properties of these exoplanets as a function of phase via phase-resolved spectroscopy in the 1-1.65 $\mu$m range for a handful of Hot Jupiters in the 1-1.6 $\mu$m range with \textit{HST}/WFC3 \citep[see, e.g., ][]{kreidberg:2018, arcangeli:2019, arcangeli:2021, jacobs:2025}. Given the anticipated level of stability of the \textit{JWST} observatory and its wide wavelength coverage, the expectations were high for the observatory to significantly enlarge both the targets on which to perform studies like those, but also the discovery space to constrain the multi-dimensional properties of exoplanets. A little over two and a half years after initiating scientific operations, we can confidently say \textit{JWST} has more than delivered on its promise to characterize exoplanets in multiple dimensions. We illustrate some of the highlights in this ongoing multi-dimensional gas giant exoplanet exploration with \textit{JWST} to date in Figure \ref{fig:5}. 

The very first showcase of the capabilities of the observatory for this multi-dimensional exploration came from the \hbindex{phase-curve} of the ultra-hot (2400 K), Jupiter-mass exoplanet WASP-121~b between 3-5 $\mu$m presented in \citet{mikal-evans:2023}. One of the most remarkable features of the band-integrated lightcurves presented in that work --- one of which we showcase in Figure \ref{fig:5}, bottom panel --- is the fact that they have \textit{not} been corrected for instrumental systematics, showcasing the superb stability of the instrumentation onboard \textit{JWST} --- even a hint of nightside emission can be spotted by eye directly in this lightcurve. The data is so precise, and taken at such high relative spectral resolution (for space-based instrumentation standards --- $R \sim 3,000 $), that the \textit{\hbindex{radial-velocity} of the planet} can be extracted directly from this data, providing the very first time an entire radial-velocity curve has been measured during its entire orbit. This data, in turn, provides hints of winds on the exoplanet through a tentative observed difference in the radial-velocity semi-amplitude at different wavelengths \citep{sing:2024-w121}. 

To date, similarly striking \textit{JWST} phase-curves have been presented in the literature for other exoplanets. One of them is the one for the highly eccentric ($e=0.93$), 4 Jupiter-mass exoplanet HD 80606~b, whose observed NIRSpec phase-curve shows hints of spectroscopic variations (i.e., ``seasonal changes") as a function of phase \citep{sikora:2024}. Another is the NIRISS/SOSS phase-curve of the hot (2000 K), 1.7 Neptune-mass LTT 9779~b \citep{coulombe:2025}, which showcases constraints on the distribution of reflected light throughout its orbit around the star, showing hints of an asymmetric dayside. We also highlight the MIRI/LRS phase-curve of the hot (1400 K), 2 Jupiter-mass exoplanet WASP-43~b studied in detail in \cite{bell:2024}, which we highlight in Figure \ref{fig:5} (top right panel). This work --- part of the ERS transiting exoplanet team observations --- reveals a stunning set of emission spectra as a function of phase, which shows strong water features at all phases, significant flux in the nightside (phase 0) believed to be coming from clouds, and a significantly higher flux when the evening is facing the observer (phase 0.25) than when the morning is facing the observer (phase 0.75). While CH$_4$ was expected to be present in the nightside, no evidence of this molecule is observed which suggest \hbindex{disequilibrium chemistry} processes are at place in the nightside of WASP-43~b. This is the first time we are observing such detailed, phase-resolved emission for an exoplanet in the 5-10 $\mu$m range.

The second set of studies we highlight are those employing the ``\hbindex{transit mapping}" technique --- the extraction of the morning (night to day) and evening (day to night) exoplanet terminator spectra from an exoplanet transit lightcurve as a function of wavelength \citep{vonparis:2016, powell:2019, espinoza:2021}. While predictions that the morning and evening limb spectra should have distinct spectroscopic signatures were predicted more than a decade ago \citep{fortney:2010}, the direct detection and study of the effect at infrared wavelengths ---key to constrain bulk properties of those limbs, such as C/O ratios, metallicites, cloud properties--- was not possible until the advent of \textit{JWST}'s unique wavelength coverage and stability. The first detection of this effect with \textit{JWST} was presented in \citet[][Figure \ref{fig:5}, middle top panel]{espinoza:2024} for the NIRSpec/PRISM observations of the exoplanet WASP-39~b. This study found that the atmosphere in the evening terminator seems to be warmer than the morning terminator by about 100 K --- a result that is in line with predictions from GCMs. An application of this very same technique on the \textit{JWST} NIRCam/Grism observations of WASP-107~b by \citet{murphy:2024} found evidence of morning and evening spectral differences as well, finding hotter evenings than mornings. These studies mark the first time we are able to resolve the spectra of the morning and evening terminators of exoplanets with low resolution transmission spectroscopy, techniques that will likely pave the way for future studies on other exoplanets to explore how these vary with different bulk properties.

Finally, we highlight the work of \citet[][Figure \ref{fig:5}, top left panel]{hammond:2024}, which produced for the first time the 2-dimensional map of an exoplanet via ``\hbindex{eclipse mapping}" studies of WASP-43~b using MIRI/LRS observations. The eclipse mapping technique \citep{rauscher:2007,cowan:2018} takes advantage of the fact that during an eclipse, different parts of an exoplanet's dayside eclipse the star at different times during ingress and egress --- which allows one to map the surface flux map of the object. Traditionally, this technique ---together with phase-curves--- has allowed to mainly constrain longitudinal information \citep[see, e.g.,][]{cowan:2013, challener:2023}. WASP-43~b, however, has a favorable ``stellar edge" angle \citep{boone:2024}, and passes behind the star in eclipse at an angle that makes contributions to the mapping signal come from both, longitudinal and latitudinal information --- the exact information the work of \cite{hammond:2024} takes advantage of to create the first longitudinal and latitudinal ---i.e., a 2D--- map of an exoplanet. This enables to put strong constraints on possible predicted latitudinal asymmetries, which have been suggested to be produced by atmospheric dynamics or even magnetic fields. Eclipse map studies for this and other exoplanets are being generated by \textit{JWST} at this and other wavelengths, and are providing unprescedented details on the thermal map structure of exoplanets \citep[see, e.g.,][]{coulombe:2023, valentine:2024, challener:2024}.

\subsection{New giant exoplanet candidates from \textit{JWST} high-contrast imaging efforts}

\begin{figure}
\includegraphics[scale=0.45]{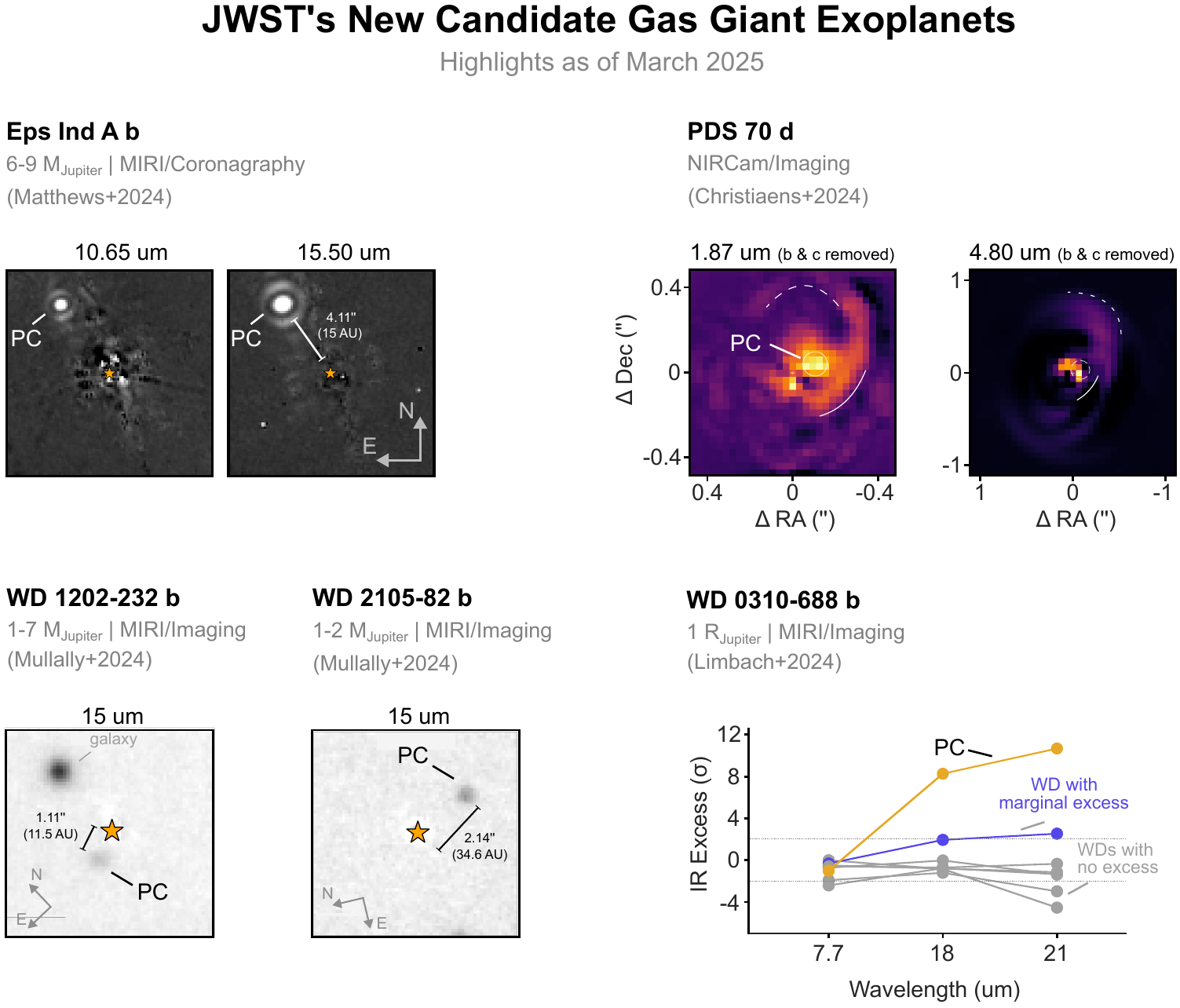}
\caption{\textbf{Some highlights on new (exo)planet candidates (PC) or exploration of already existing exoplanet candidates with \textit{JWST}.} (Top left) Identification of the exoplanet candidate Eps Ind A~b, a cold (275 K) super-Jupiter --- \textit{JWST}'s first exoplanet detection with MIRI/coronography, after subtracting the star (orange in the center) \citep[adapted from][under a Creative Commons Attribution (CC BY) license]{matthews:2024}. (Top right) PDS 70 observations with NIRcam imaging after subtracting the star, disk and exoplanets PDS 70~b and c from the image. A residual signal is observed, which is consistent with the location of exoplanet candidate PDS 70 d, identified from the ground \citep[adapted from][under a Creative Commons Attribution (CC BY 4.0) license]{christiaens:2024}. (Bottom left) Two Jupiter-mass exoplanets identified using MIRI photometry orbiting two different white dwarfs \citep[adapted from][]{mullally:2024}. (Bottom right) A Jupiter-size, cold (250 K) exoplanet candidate identified on the white dwarf WD 0310-688 via multi-band infrared excess \citep[orange curve; adapted from][]{limbach:2024}.}
\label{fig:6}       
\end{figure}

One of the revolutionary aspects of the instrumentation onboard \textit{JWST} for exoplanet science is its stability and sensitivity at long wavelengths, which allows the observatory to detect signals of exoplanets on a wide range of environments, from currently forming exoplanets to cold exoplanets orbiting dead stars. This makes it a unique laboratory to detect exoplanets at long distances at or below the masses of Jupiter if they are there ---- and to tightly constrain formation mechanisms and processes if they are not  \citep[see, e.g.,][and references therein]{carter:2021}.

One of the first highlights in the delivery of new \hbindex{\textit{JWST} directly imaged exoplanets} came from the identification of two new exoplanet candidates orbiting the \hbindex{white dwarfs} WD 1202-232 and WD 2105-82 via MIRI/imaging at 15 $\mu$m by \cite[Figure \ref{fig:6}, bottom left panel; from][]{mullally:2024}. These discoveries not only marked the first set of giant exoplanet candidate identifications from \textit{JWST} but, if confirmed, these would mark the first time we are able to detect exoplanets at distances similar to the planets in our own Solar System, around stars of ages similar or older than our own Sun and at masses similar to our own Solar System gas giants. 

Another highlight in the search for exoplanets orbiting white dwarfs is that of the search being performed by the MIRI Exoplanets Orbiting White dwarfs (MEOW) Survey \citep{limbach:2024}. This survey is using the \hbindex{infrared excess} technique on the search of exoplanets, which consists on comparing multi-band photometry at long wavelengths with model spectral energy distributions (SEDs) of stars: when observations reveal larger fluxes than the stellar SEDs, then there is evidence for an exoplanet. With \textit{JWST}, this technique is particularly powerful, allowing in principle the detection of exoplanets all the way down to the size of the Earth around nearby white dwarfs \citep{limbach:2022}. The work of \cite{limbach:2024} reveals the first solid candidate from this survey: WD 0310-688~b (see Figure \ref{fig:6}, bottom right panel). The infrared excess was used in that work to estimate that it is consistent with a Jupiter-size, cold (250 K) exoplanet. The closeness of the system to the Earth (only 10 parsecs) would make this an exciting object if confirmed, allowing a wide range of follow-up spectroscopic characterization possible.

\textit{JWST} is also making clear advances in detecting signals of exoplanets at the very beggining of their lives, all the way to exoplanets around main-sequence stars. Among the highlights on this search is the identification of Eps Ind A~b --- a 6-9 Jupiter-mass, cold (275 K) exoplanet candidate orbiting a main sequence K star \citep[Figure \ref{fig:6}, top left panel][]{matthews:2024}, which marks \textit{JWST}'s first exoplanet orbiting a main sequence star --- and the first direct image exoplanet orbiting a star with an age comparable to our own Sun. Another highlight of this \textit{JWST} search for exoplanets involves observations of the PDS 70 system --- which has two known directly imaged, still forming exoplanets --- by both \citet{christiaens:2024}, using NIRCam imaging (see Figure \ref{fig:6}, right panel) and \citet{blakely:2024}, using NIRISS/AMI. These works were able to put new strong constraints not only on possible circumplanetary disks around the PDS 70~b and PDS~70~c exoplanets, but were also able to constrain fluxes of the possible planet PDS 70~d, identified from ground-based observations by \citet{mesa:2019}.

Finally, we highlight the recent announcement of a planetary candidate in the TWA 7 system \citep{lagrange:2025}. Detected via \textit{JWST} MIRI, this exoplanet, if confirmed, would be unique in many ways. First, it would serve as one of the best systems to fully characterize planet-disk interactions, as the estimated planetary mass and distance could explain the main disk structures observed on this system. Second, its low estimated mass (0.3 M$_J$), would earn the planet the record for the lowest mass exoplanet ever imaged --- opening the window to perform detailed characterization of sub-Jupiter mass exoplanets for directly imaged exoplanets.


\section{Highlights of \textit{JWST} observations of sub-Neptunes}

One of the most important discoveries of the \textit{Kepler} mission \citep{kepler:2010} was to find that the most common type of planet in our galaxy, at least in orbits with periods smaller than about 100 days, are those with sizes between 1$R_\oplus$ and 4$R_\oplus$ \citep[see, e.g., ][for a review]{bean:2021}. These small exoplanets seem to come, in turn, predominantly in two sizes: planets smaller than about 1.5$R_\oplus$ are believed to be rocky ``super-Earths", while planets larger than about 2$R_\oplus$ are believed to have significant volatile and/or H/He contents \citep{fulton:2017, fulton:2018}. Planets between 1.5-2$R_\oplus$, on the other hand, reside in a radius "gap" in the occurrence rate distribution of these small exoplanets, some of which are believed to be primarily enriched in volatiles \citep[e.g., water worlds;][]{luque:2022}. In this Section ---motivated in large part by recent \textit{JWST} discoveries and insights we discuss below--- we use the term ``\hbindex{sub-Neptune}" to refer to any exoplanet with a radius smaller than about 4$R_\oplus$ that \textit{might}, either through density or atmospheric constraints, have a significant H/He and/or volatile envelope (e.g., an envelope rich in H$_2$O, CO$_2$, CO, etc.). We refer to exoplanets \textit{without} such a volatile envelope as ``rocky" exoplanets --- very similar to the definition employed in the work of \cite{rogers:2015}. A discussion of JWST highlights in rocky exoplanet exploration (including studies of super-Earths, the sub-Neptune counterparts in our nomenclature) is presented in the next Section.

The sub-Neptune exoplanet population is as enigmatic as it gets. Given no such planet has been detected to date in our Solar System \citep{p9:2021, siraj:2023} they defy our intuition in terms of predicting what they should look like, what they are made of and how they form and evolve \citep{rogers:2021,schlichting:2022, misener:2023, seo:2024}. While precise constraints on mass and radius for this population has enabled significant advances in our understanding of their possible compositions \citep[through densities and interior modeling, see, e.g.,][]{lopez:2014, rogers:2015, dorn:2017, luque:2022, misener:2023}, the ultimate test to these predictions relies on precisely characterizing their atmospheres. While \textit{HST} in particular made tremendous progress in delivering the first hints on what those might look like \citep[see, e.g.,][]{kreidberg:2014,knutson:2014,bourrier:2017,tsiaras:2019,benneke:2019, evans:2023}, definite conclusions on the actual compositions these hints implied could only be made with \textit{JWST} observations \citep[see, e.g., the case of K2-18~b;][]{madhusudhan:2020,barclay:2021,bezard:2022}. Atmospheric characterization of sub-Neptunes, thus, was a highly anticipated outcome of the \textit{JWST} mission. Here, we highlight some of the initial results in the past few years that are leading the way in this exciting exploration of their atmospheres and interiors.

\subsection{Introducing sub-Neptune atmospheric exploration with GJ 1214~b}

The sub-Neptune \hbindex{GJ 1214~b} \citep{charbonneau:2009} is perhaps one of the finest examples of the complexity of studying the atmospheres of sub-Neptunes and thus the perfect example to highlight the role \textit{JWST} is having in exploring these distant worlds. This 8.4$M_\oplus$, 2.7$R_\oplus$ relatively warm (560 K) sub-Neptune had, for many years, one of the most precise transmission spectra ever obtained for an exoplanet: the one presented in \citet{kreidberg:2014}, and obtained via 15 transits of the exoplanet observed by \textit{HST}/WFC3. This extreme precision data revealed a featureless transmission spectrum, with the most likely explanation being that clouds and/or a high metallicity atmosphere were damping all of its molecular features observable in transmission. This degeneracy between clouds and metalliciy \citep[see, e.g.,][]{line-parmentier:2016} made it impossible to conclusively discern with this data alone whether the atmosphere was primarily primordial (e.g., made of H/He) with high altitude clouds obscuring any spectroscopic features, or enhanced in metals with clouds. Furthermore, detailed studies showed that a wide range of clouds or hazes could give rise to the observed spectrum, with long-wavelength observations such as the ones achievable with \textit{JWST} being critical to constrain cloud/haze properties and help break the cloud-metallically degeneracy \citep[see, e.g.,][]{morley:2013, barstow:2013, kataria:2014, charnay:2015, gao:2018, kawashima:2019, lavvas:2019, christie:2022}. 

\begin{figure}
\includegraphics[scale=0.45]{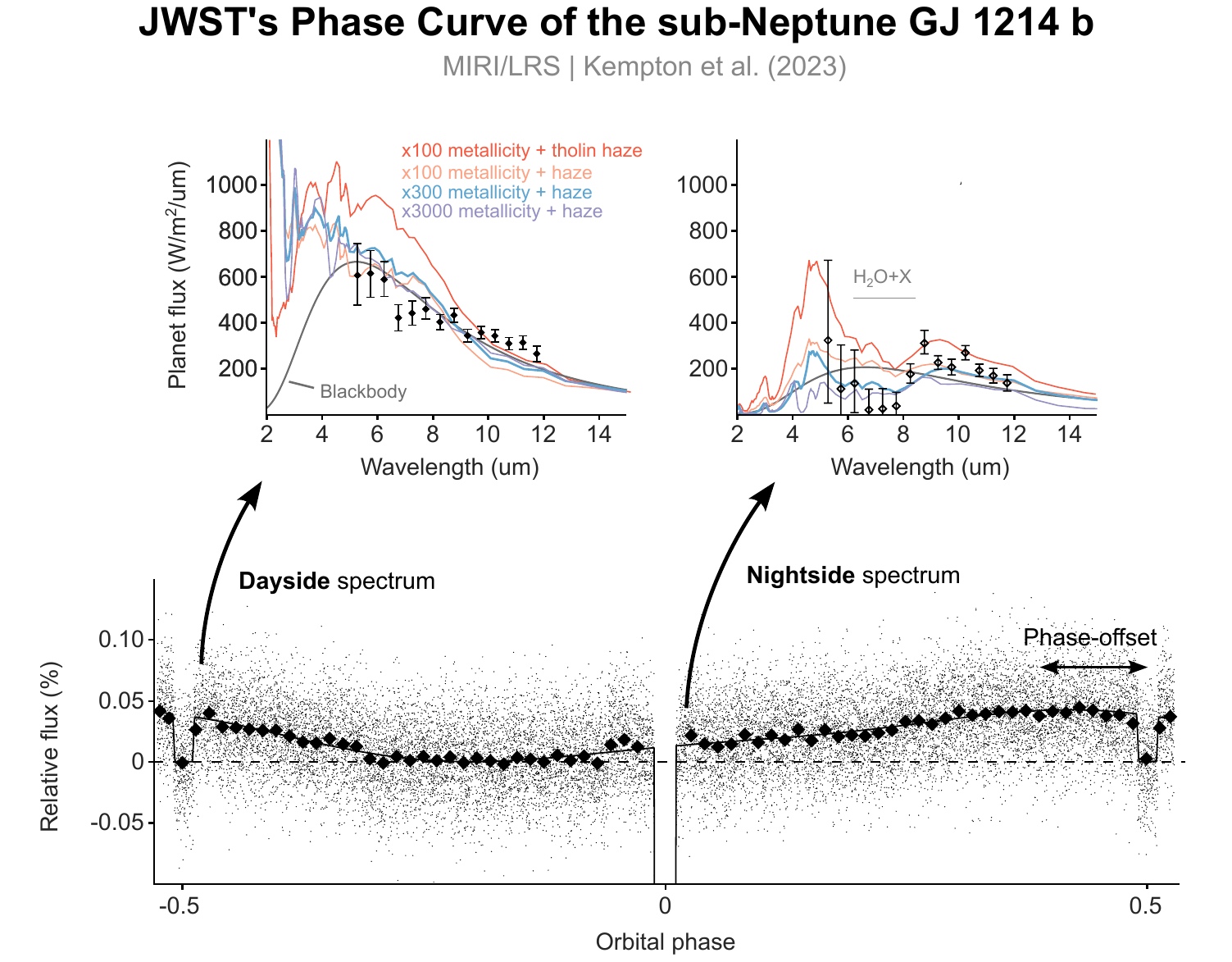}
\caption{\textbf{The phase-curve of the sub-Neptune GJ 1214~b as observed by \textit{JWST}.} (\textit{Bottom}) Band-integrated (i.e., white-light) light curve of the phase-curve of GJ 1214~b as observed by MIRI/LRS. Note the maximum of the phase curve does not occur exactly at secondary eclipse, but shows a ``phase-offset" shift. (\textit{Top}) Dayside and nightside spectrum of GJ 1214~b as derived from the phase-curve, along with high-metallicity GCM models with highly reflective hazes (purple, blue and orange curves), tholin haze (red curves) and blackbodies (grey curves). Note how the nightside in particular deviates from a blackbody, with a dip at about 7 $\mu$m which is consistent with being absorption of H$_2$O and either CH$_4$ or HCN (X, in the annotation). Adapted from \cite{kempton:2023}.}
\label{fig:7}       
\end{figure}

In Figure \ref{fig:7}, we showcase the stunning \textit{JWST} MIRI/LRS \hbindex{phase-curve} of GJ 1214~b together with the day and nightside phase-resolved emission spectra obtained and interpreted by the team of \cite{kempton:2023}. This work confirms that, indeed, the atmosphere has to have a high metallicity and include clouds or hazes at all phases. Interestingly, exploration of the phase-resolved emission spectra gives rise to several novel insights into GJ 1214~b's atmospheric make-up. First, the observed brightness of the planet inferred from the phase-curve seems to be much smaller than expected, implying not all input energy from the star is making it into the planet ---  using these \textit{JWST} observations, \cite{kempton:2023} derives a relatively large ($A_B\sim 0.5$) albedo, providing strong evidence of it having highly reflective clouds/hazes. Second, by comparing GCM modelling to the observed spectra, a range of possible hazes were able to be ruled out (both sooth and tholin hazes; see, e.g., red curve in Figure \ref{fig:7});  these comparisons, in turn, require high metallicity atmospheres to explain the observed day-to-night flux contrast, with the nightside spectrum even showing an apparent absorption feature at about 7 $\mu$m which can be interpreted as arising from H$_2$O and either CH$_4$ or HCN. These varied level of constraints highlight the ability of such observations to unveil cloud/haze properties and metallicities down to sub-Neptune class exoplanets through phase-resolved emission spectroscopy with \textit{JWST}.

Another piece of the puzzle of GJ 1214~b's atmospheric composition was recently added via transmission spectroscopy with \textit{JWST} by the work of \cite{ohno:2025}. This work jointly studies the \textit{HST}/WFC3 observations of \cite{kreidberg:2014}, the \textit{JWST} MIRI/LRS transmission spectrum obtained during the phase-curve observations of \cite{kempton:2023} and the \textit{JWST} NIRSpec/G395H transmission spectrum of GJ 1214~b obtained by \citet{schlawin:2024} which showcases tentative evidence for CO$_2$ in its transmission spectrum (see Figure \ref{fig:8}). Under a wide variety of assumptions, \cite{ohno:2025} derives an atmospheric metallicity in excess of $\times$1000 solar --- in other words, a metal-dominated atmosphere.  Such a result would put this exoplanet atmosphere ---when paired with its possible internal structure in light of these \textit{JWST} observations--- outside the realm of our \hbindex{Solar System} intuition. Among the possibilities for GJ 1214~b is a rocky exoplanet-like atmosphere (i.e., high mean molecular weight) on top of a Neptune-like interior --- with corresponding volatile envelopes possibly in excess of 50\% for this level of metallicity \citep{nixon:2024}. As we touch on the next sub-section, this intuition-defying trend is an emerging feature of the sub-Neptune exoplanetary science that \textit{JWST} is starting to unveil.

\subsection{JWST's exploration of Sub-Neptune exoplanet atmospheres}

As demonstrated by the case study of GJ 1214~b in the previous subsection, the atmospheric exploration of sub-Neptune exoplanets with \textit{JWST} is a highlight on itself. These observations, in turn, seem to be solidifying the view that sub-Neptune exoplanets might have distinct structures to those observed in our Solar System planets, a view that has been widely discussed and proposed by theoretical investigations in the literature in the past decade \citep[see, e.g.,][]{moses:2013, hu:2014, zeng:2019, venturini:2020, guzman:2022}. \textit{JWST} exoplanet programs are exploring the diversity of sub-Neptune exoplanet atmospheres via two different strategies: either performing deep exploration of a few special sub-Neptunes, e.g., GJ 1214~b \citep{kempton:2023, schlawin:2024}, K2-18~b \citep{madhusudhan:2023, hu-prop:2021}, TOI-270~d \citep{benneke:2024, holmberg:2024}, LHS 1140~b \citep{damiano:2024, cadieux:2024} --- which we discuss below --- or by performing wide surveys of small exoplanets to unveil statistical trends, with the finest example being the COMPASS (Compositions of Mini-Planet Atmospheres for Statistical Study) program of \citet[][PID 2512]{compass:2021}. Here, we opt to center our discussion around the deep dives performed in a handful of sub-Neptunes, as statistical trend results on sub-Neptunes using \textit{JWST} data have not been published to our knowledge at the date of writing.

\begin{figure}
\includegraphics[scale=0.45]{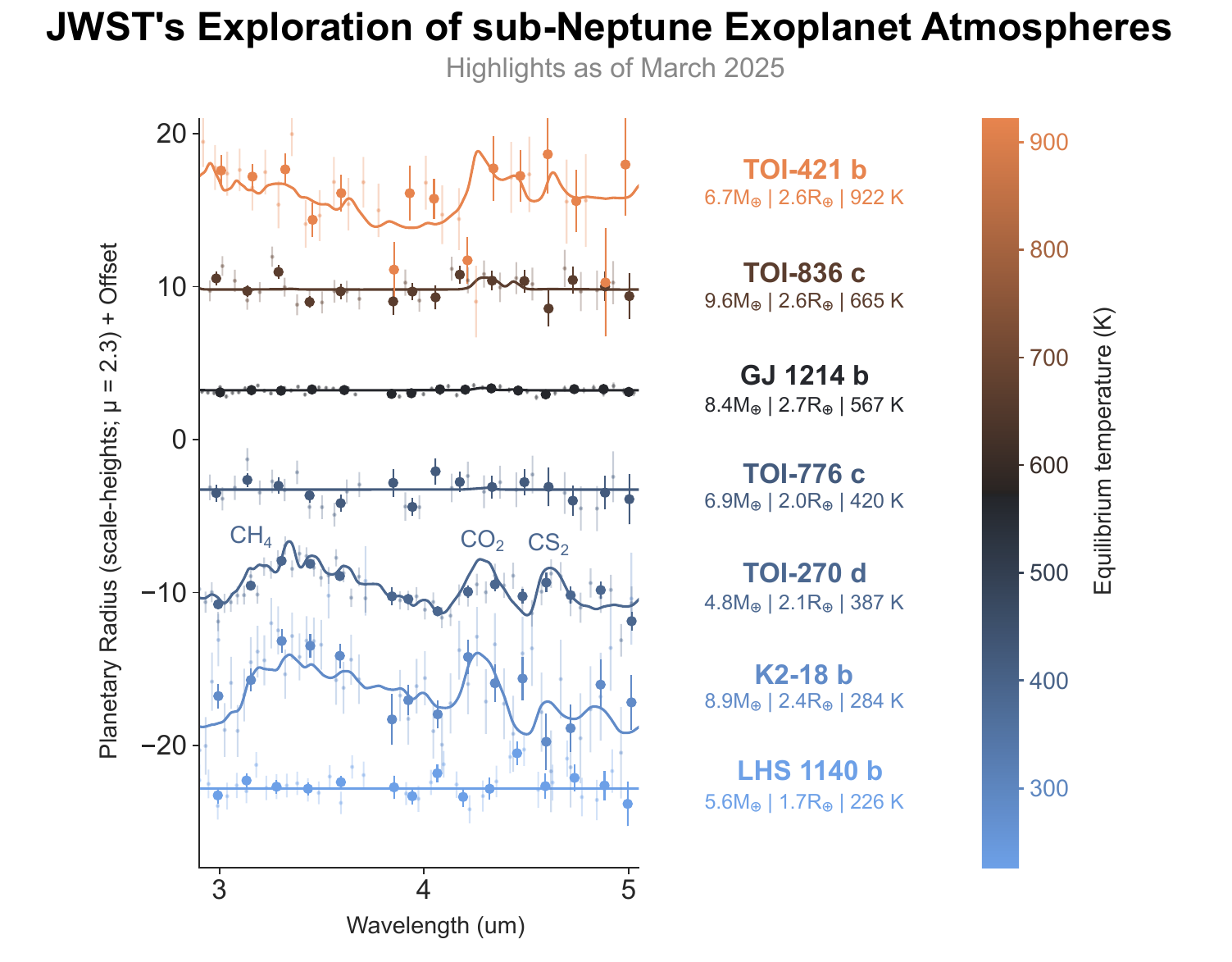}
\caption{\textbf{Highlights of \textit{JWST} exploration of sub-Neptune exoplanet atmospheres via transmission spectroscopy.} Transmission spectra of sub-Neptunes with temperatures from 220 K (blue) to 670 K (orange), obtained by \textit{JWST} NIRSpec/G395H in the 3-5 $\mu$m range, showcasing the diversity on the composition and observable features in their atmospheres (CH$_4$ at 3.5 $\mu$m, CO$_2$ at 4.3 $\mu$m, CS$_2$ at about 4.6 $\mu$m). The y-axis is the planetary radius normalized by the atmospheric scale-height of each planet assuming a H/He mixture ($\mu = 2.3$; likely \textit{not} the mean molecular weight in these atmospheres, only useful for normalization --- see text for details). This is offset for each planet for clarity. From top to bottom, transmission spectrum of TOI-421~b \citep[from][]{davenport:2025}, TOI-836~c \citep[from][]{wallack:2024}, GJ 1214~b \citep[from][]{schlawin:2024}, TOI-776~c \citep[from][]{teske:2025}, TOI-270~d \citep[from][]{holmberg:2024}, K2-18~b \citep[from][]{madhusudhan:2023} and LHS 1140~b \citep[from][]{damiano:2024}. Solid lines show H$_2$-dominated, cloudy, best-fit POSEIDON \citep{poseidon1, poseidon2} forward models; they only include CH$_4$, CO$_2$ and CS$_2$ as chemically active species; shown for illustration purposes only.}
\label{fig:8}       
\end{figure}

To highlight the current status on the exploration of sub-Neptune exoplanets by \textit{JWST}, in Figure \ref{fig:8} we showcase publicly available \textit{JWST} transmission spectra obtained in the 3-5 $\mu$m range via NIRSpec/G395H ---a particularly important wavelength range to search for features of H$_2$O, CH$_4$, CO$_2$ and other molecules--- for 7 relatively warm ($< 1000$ K) sub-Neptunes. In this illustration, we have decided to present the planetary radius as a function of wavelength, but normalized by the corresponding scale-height for each planet assuming a H/He-dominated mixture. While this is very likely an invalid assumption for many (or even all) of the planets in this sample ---notably for GJ 1214~b, as discussed in the previous sub section--- this allows us to put all the spectra on the same relative scale, which we find useful to highlight some very interesting trends. 

First, note how GJ 1214~b, despite having a similar mass and radius to K2-18~b, has a much more compact atmosphere. As we discuss below, this is very likely due to their very different metallicities ($>\times 1000$ solar for GJ 1214~b, of order $\times 100$ solar for K2-18~b), which imply very different mean molecular weights --- and thus atmospheric make-ups. Relatedly, it does not seem there is a clear pattern for predicting which sub-Neptunes will give rise to large amplitude features in their transmission spectrum at this wavelength range. This might once again hint towards chemical diversity in their atmospheres. Arguably, one of the most striking set of spectra showcased in this illustration are those of the cooler K2-18~b and TOI-270~d, as these reveal large spectroscopic features. Despite LHS 1140~b having similar mass, radius and temperature to TOI-270~d, it shows a much flatter overall spectrum in this wavelength range. This makes it an interesting case study as well, for similar reasons to why GJ 1214~b seems to be relatively compact. We highlight and discuss findings enabled by \textit{JWST} transmission spectroscopy studies of K2-18~b, TOI-270~d and LHS 1140~b below, as they have been triggering an important number of studies in the literature, including follow-up observations with \textit{JWST} itself.

\subsubsection{K2-18~b: a hycean world, or mini-Neptune?}

While similar in mass and radius, \hbindex{K2-18~b} and GJ 1214~b have one key distinct feature: the former is much warmer, with an equilibrium temperature (280 K) reminiscent of that of Earth. The transmission spectrum showcased in Figure \ref{fig:8}, presented in \citet[][which includes in addition 0.7-3 $\mu$m NIRISS/SOSS observations not shown here]{madhusudhan:2023}, highlights a solid detection of CH$_4$ in its atmosphere \cite[predicted to be the dominant absorber based on \textit{HST} observations by][]{bezard:2022}, and hints of ---or at least upper limits on--- other molecules, such as CO$_2$ and even \hbindex{dimethyl sulfide} \citep[DMS --- a potential biomarker; but see][]{schmidt:2025}. What is more, this study provides \textit{abundance constraints} on those molecules, which has initiated a wide range of theoretical investigations to unveil possible interiors for K2-18~b \citep{wogan:2024, tsai:2024, cooke:2024, schmidt:2025}. These range from the exoplanet being a miniaturized version of Neptune, to the possibility of habitable conditions on this sub-Neptune via, e.g., hycean scenarios --- a \textit{hy}drogen-dominated atmosphere sitting on top of a liquid water o\textit{cean} \citep[see, e.g.,][]{hycean1, hycean2}. A thorough assessment of those possibilities, along with a summary of the relevant literature is discussed at length in \cite{schmidt:2025}. At the time of writing, it is safe to say the jury is still out on the true nature of K2-18~b's atmosphere and interior. A much more precise transmission spectrum which could unveil this answer is imminent as the \textit{JWST} program of \citet[][PID 2372]{hu-prop:2021} is obtaining a spectrum that should be at least twice as precise as the one obtained by \cite{madhusudhan:2023}.

\subsubsection{TOI-270~d: a likely miscible-envelope sub-Neptune}

We now turn our discussion to the equally exciting sub-Neptune \hbindex{TOI-270~d}, which has about the same size but half the mass (4.8M$_\oplus$) of K2-18~b, is relatively warm (387 K), and whose \textit{JWST} transmission spectrum was presented both in \citet[][which includes additional NIRISS/SOSS data, not shown here]{benneke:2024} and in \cite{holmberg:2024}. As illustrated in Figure \ref{fig:8}, this transmission spectrum is feature rich, showing solid detections of CH$_4$, CO$_2$ and even hints of \hbindex{carbon disulfide} \citep[CS$_2$, a potential biomarker;][]{seager:2013, hycean2} in its atmosphere. Both studies constrain the atmosphere to be significantly enriched in hydrogen, but the study of \cite{benneke:2024} provides many interesting insights into the nature of TOI-270~d. First, using these \textit{JWST} observations, this works constrains the mean molecular weight of this sub-Neptune to be $5.47^{+1.25}_{-1.14}$, significantly larger than Neptune/Uranus ($\mu = 2.6$) and thus making it unlikely to contain \textit{only} H/He. This constraint, in turn, suggests that TOI-270~d doesn't follow the structures discussed for other sub-Neptunes such as, e.g., K2-18~b, whose more massive nature allows it to retain a thin H/He envelope. Instead, TOI-270~d might be a ``miscible-envelope sub-Neptune" --- an exoplanet whose envelope and atmosphere is a well mixed combination of H$_2$ and high mean molecular weight volatiles (e.g., H$_2$O). The second set of insights the work of \cite{benneke:2024} makes is in relation to TOI-270~d's interior. Using these atmospheric constraints, together with its mass and radius, this work estimates that the structure of the exoplanet must have about 90\% of its mass (4.3M$_\oplus$) in rock/iron, with about 6\%  (0.3M$_\oplus$) in volatiles and 4\% (0.2M$_\oplus$) in H/He. This suggests TOI-270~d might be more connected to rocky exoplanets in our own Solar System than previously thought --- being able to retain its volatile envelope only because of its larger mass, a fate our lower mass Solar System rocky exoplanets were not able to follow.

\subsubsection{LHS 1140~b: a likely high mean molecular weight atmosphere in a habitable-zone exoplanet}

We conclude this section on the highlights of the exploration of sub-Neptune exoplanets by \textit{JWST} with a very special member of this exoplanet family: \hbindex{LHS 1140~b}. This 1.7$R_\oplus$, 5.6$M_\oplus$ exoplanet has gained attention in the exoplanet community due to both its temperate nature, which puts it in the \hbindex{habitable-zone} of its star, and due to its remarkably precise bulk properties which, as TOI-270~d, allow for compositions that might not follow the ones we are used to in our Solar System planets \citep{cadieux-mass:2024}. Among the possible scenarios, these interior modeling efforts allow for a miniaturized version of Neptune, with a massive ---likely unsuitable for life--- H$_2$-dominated atmosphere, a water world; a volatile-rich envelope with a relatively high mean-molecular weight atmosphere on top which could be potentially habitable, or even a hycean world. Interestingly, independent \textit{JWST} transmission spectroscopy studies by \cite{damiano:2024}, who used both NIRSpec/G395H (shown in Figure \ref{fig:8}) and NIRSpec/G235H (not shown here) and \cite{cadieux:2024}, who used NIRISS/SOSS (not shown here), both reach the same conclusion: this exoplanet is unlikely to have a massive H$_2$-dominated atmosphere \citep[but see][]{huang:2024}. These studies, thus, strongly suggests LHS 1140~b could have a high mean molecular weight atmosphere on top of a volatile envelope, or even a mixed rock/volatile interior. The only way to constrain whether LHS 1140~b could have habitable environments for life via such an atmosphere is through future \textit{JWST} observations aimed at detecting spectroscopic features in the atmosphere. 6 more transits have indeed been selected to be observed in Cycle 4 to further characterize this exciting exoplanet (PID 7073, co-PIs: Lustig-Yaeger \& Stevenson). 

\section{Highlights of \textit{JWST} observations of rocky exoplanets}

From Mars and Venus' CO$_2$-dominated atmospheres, to Earth and Titan's N$_2$-dominated atmospheres, the rocky planets and satellites in our Solar System show a wide variety of atmospheric compositions that to this day we are sill trying to decipher and make sense of through remote and in-situ measurements \citep[see, e.g.,][for reviews]{horst:2017, taylor:2018, vandaele:2024}. Whether rocky exoplanet atmospheres in stellar systems elsewhere are also similarly diverse ---and in particular whether some of the more close-in ones could even hold on to atmospheres--- is today one of the most fundamental questions in exoplanetary science \citep{astro2020, redfield:2024}. \textit{JWST} provides our very first window into exploring answers to this question in detail. 

Prior observations of a handful of rocky exoplanet systems did set the stage for this \textit{JWST} exploration. For example, initial \textit{HST}/WFC3 reconnaissance observations of the exoplanets in the \hbindex{TRAPPIST-1} system via transmission spectroscopy enabled to constrain not only the absence of cloud-free, H$_2$-dominated atmospheres in most of the planets \citep[see, e.g.,][]{dewit:2018, garcia:2022, gressier:2022}, but also to identify the \hbindex{transit light source effect} \citep[TLS;][]{rackham:2018} as one of the main challenges when interpreting such observations \citep{zhang:2018}. Similar results were obtained for the rocky exoplanets GJ 1132~b,  L 98-59 b, c and d \citep{lr:2022, damiano:2022, zhou:2022, barclay:2023, zhou:2023}, with somewhat stronger constraints on LTT 1445 A~b \citep{bennet:2025} --- all spectra being at various levels consistent with being featureless. The \textit{Spitzer} space telescope was also able to explore the nature of hot, ultra-short period ($P<1$ day) rocky exoplanets. The now famous 4.5 $\mu$m phase-curve of LHS 3844~b presented in \cite{kreidberg:2019} conclusively demonstrated this exoplanet lacked a thick ($>$10 bar) atmosphere. The similarly iconic \textit{Spitzer} phase-curve of 55 Cancri~e introduced by \cite{demory:2016} provided, in turn, the first hints of an exoplanet atmosphere on a rocky exoplanet, a suggestion that has stood the test of time through several re-analyses \citep[see, e.g.,][]{mercier:2022}. In a similar vein, the secondary eclipse and corresponding \textit{Spitzer} phase-curve of K2-141~b also provided some evidence for a rock vapor atmosphere in this hot, rocky super-Earth \citep{zieba:2022}.

With its large diameter aperture and wide wavelength range coverage, \textit{JWST} is a unique observatory to explore rocky exoplanet atmospheres. On the one hand, it covers wavelengths on which various important absorbers such as CO$_2$ ---ubiquitous throughout the rocky planets in our Solar System--- have, in principle, detectable spectroscopic features \citep[see, e.g.,][]{turbet:2020,wk:2022}. Its infrared capabilities also allow it to detect thermal emission from rocky exoplanets elsewhere in order to measure the planet's energy budgets, and even help differentiate through such measurements between bare rocks and objects with atmospheres in particular for exoplanets orbiting M-dwarfs \citep[see, e.g.,][]{koll:2019, mansfield:2019}. Here, we provide some highlights stemming from the first two years of \textit{JWST}'s rocky exoplanet atmospheric exploration that make use of those strategies and techniques.

\subsection{Unveiling lava worlds with \textit{JWST}: the case of 55 Cancri~e}
\begin{figure}
\includegraphics[scale=0.45]{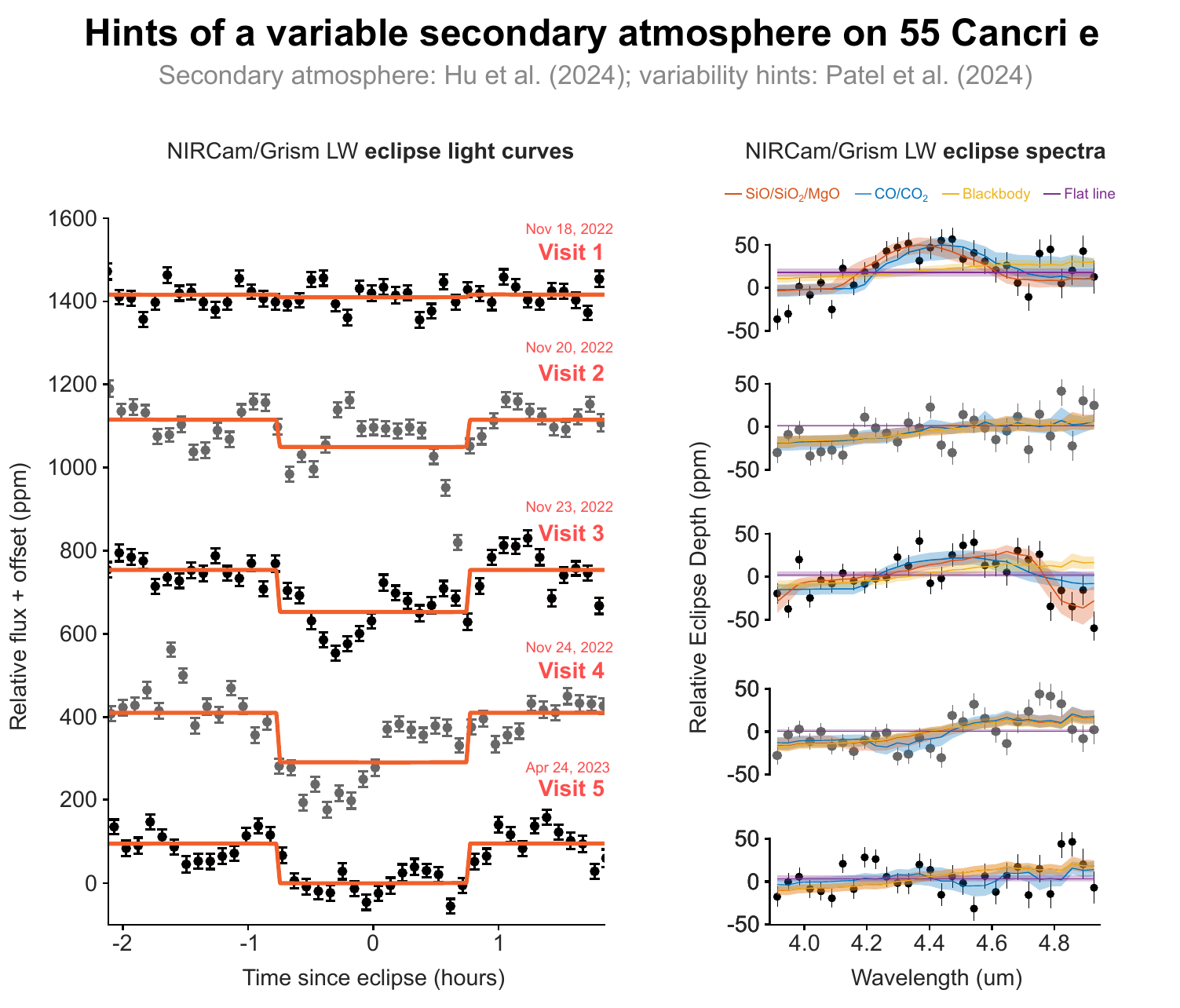}
\caption{\textbf{Evidence for a possible variable, secondary atmosphere on the lava world 55 Cancri e.} Secondary eclipse light curves (data in black, best-fit models in red --- left) and corresponding emission spectra from \textit{JWST} NIRCam's long-wavelength (LW) channel (data in black, best-fit models of different atmospheric compositions in colored solid lines, right) for 55 Cancri~e over different visits. Note how both seem to vary on few days time-scale both in the light curves and the spectra. Visit 4 is a reanalysis of the spectrum studied by \cite{hu:2024} on which the same light curve shape is observed, and a qualitatively similar spectrum was studied which provided evidence for a secondary atmosphere on 55 Cancri~e. Figure adapted from the work of \cite{patel:2024}, under a Creative Commons Attribution (CC BY 4.0) license.  Similar variability is observed in this latter work in NIRCam's 2.1 $\mu$m photometric channel; this is not shown here.}
\label{fig:9}       
\end{figure}

The first set of studies we highlight in this section are the in-depth study the works of \cite{hu:2024} and \cite{patel:2024} have performed with \textit{JWST} NIRCam and MIRI on the possibile atmospheric make-up of the hot ($T_{\textnormal{eq}} = 2000$ K) super Earth ($M_p=8.8M_\oplus$, $R_p=1.95R_\oplus$) \hbindex{55 Cancri~e}. As discussed in the introduction to this section, this exoplanet has been extensively studied in the past via \textit{Spitzer}, but observations with other telescopes also provided plenty of additional constraints on the possibility of an atmosphere on this exoplanet, which are important to understand to contextualize the \textit{JWST} observations and results. Observations of the primary transit event both from the ground \citep{jindal:2020, deibert:2021} and from space with \textit{HST}/WFC3 \citep{tsiaras:2016}, already hinted at evidence for a high mean molecular weight atmosphere on this exoplanet. The \textit{Spitzer} observations of the phase-curve and secondary eclipses, in turn, provided further evidence not only of an atmosphere \citep{demory:2016} --- but perhaps of a variable one, as the secondary eclipses observed by \textit{Spitzer} seem to vary significantly from observation to observation, an effect not observed on the transit events also observed by the telescope \citep{demory-variable:2016, tamburo:2018}. Further evidence of variability in its possible atmosphere is also showcased from MOST \citep{sulis:2019} and CHEOPS \citep{morris:2021, meier:2023, demory:2023} secondary eclipse and phase-curve observations of the exoplanet. 

In Figure \ref{fig:9}, we showcase the results obtained by \cite{patel:2024} using \textit{JWST} NIRCam/Grism observations of 55 Cancri~e through 5 secondary eclipse observations between the end of 2022 and the beggining of 2023. As can be seen both on the secondary eclipse light curve themselves and the corresponding emission spectra, in line with what is observed in previous observatories, the eclipses do show strong variation from visit to visit, even leading to a non-detection of the broadband event on Visit 1 (top light curve event, left). These \textit{JWST} observations, however, allow to constrain the \textit{abundance} of possible atmospheric features in the spectra --- a first at such long wavelengths. The detailed analysis in \cite{hu:2024}, on which the data for Visit 4 in Figure \ref{fig:9} was first showcased and analyzed, reveals that the spectrum, together with another emission spectrum obtained with the MIRI instrument by their team (not shown here), is consistent with a volatile-rich atmosphere --- likely rich in CO$_2$ or CO. \cite{patel:2024}, in addition, suggests this atmosphere is variable, as the spectrum showcases strong wavelength-dependent variations from visit to visit, which could be explained by the stochastic \hbindex{outgassing} of these molecules --- perhaps sustained by a \hbindex{magma ocean} \citep[see, e.g.,][]{heng:2023, meier:2023, loftus:2024, nicholls:2025}.

\subsection{JWST's exploration of the TRAPPIST-1 exoplanet system}

\begin{figure}
\includegraphics[scale=0.45]{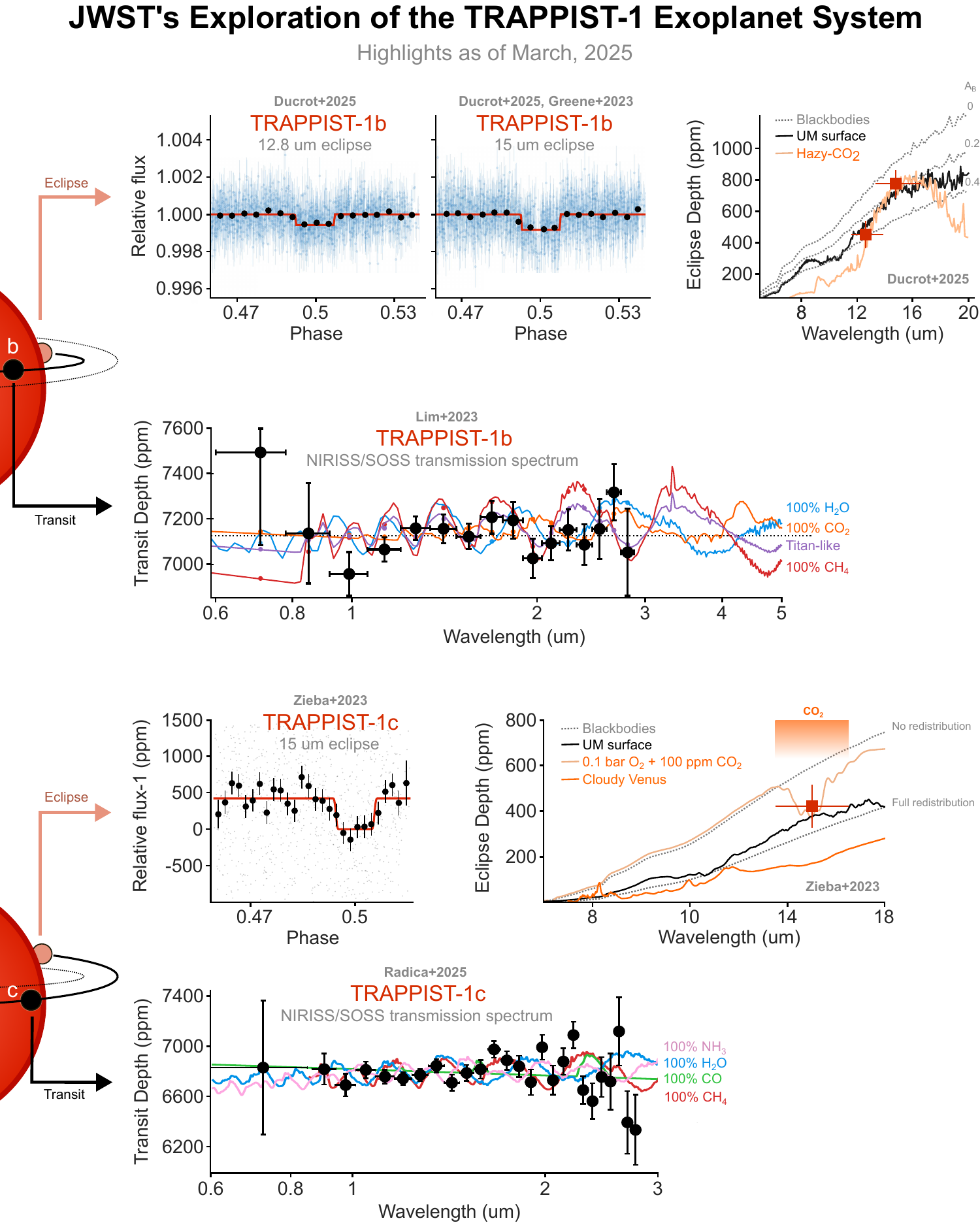}
\caption{\textbf{Highlights from \textit{JWST}'s atmospheric exploration of the rocky planets TRAPPIST-1~b and TRAPPIST-1~c.} Left-most illustration showcases the TRAPPIST-1 star (red fraction of a circle) and the orbits of TRAPPIST-1~b (inner one) and TRAPPIST-1~c (outer one), together with the geometries of the different measurements showcased. (\textit{Top}) Emission photometry and transmission spectroscopy characterization of TRAPPIST-1~b. The eclipses, and corresponding emission measurements at 12.8 and 15 $\mu$m are adapted from \citet{ducrot:2024}. Eclipse depths as a function of wavelength are showcased against blackbodies with different bond albedos $A_B$ (grey curves), an ultramafic rock surace (UM in the illustration, in black), and a hazy atmosphere that creates a thermal inversion that produces a CO$_2$ feature in \textit{emission} (orange model); also adapted from \citet{ducrot:2024} The transmission spectrum is adapted from \citep{lim:2023}, and it has been corrected for stellar contamination. The various models shown are all consistent with the data. (\textit{Bottom}) Same graphical depiction of the constraints on TRAPPIST-1~c. Eclipse and model emission spectra as a function of wavelength adapted from \cite{zieba:2023}, under a Creative Commons Attribution (CC BY 4.0) license. In eclipse, once again blackbodies (grey) are shown along an ultramafic surface (UM, black). Models for a cloudy, 10-bar Venus is shown in dark orange, thin orange curve shows a 0.1 bar O$_2$ plus 100 ppm CO$_2$ atmosphere, which is consistent with the data; also adapted from \cite{zieba:2023}, under a Creative Commons Attribution (CC BY 4.0) license. Similarly to the transmission spectrum of TRAPPIST-1~b, the spectrum of TRAPPIST-1~c is shown below, adapted from \cite{radica:2025}. Similar to TRAPPIST-1~b, the models shown (all at 100 bar) cannot be differentiated from the data.}
\label{fig:10}       
\end{figure}

The \hbindex{TRAPPIST-1} exoplanetary system \citep{guillon:2024} is one of the most studied systems by \textit{JWST}, with about 400 hours of charged telescope time devoted to studying the atmospheres of these transiting exoplanets to date. With a total of seven currently known, Earth-sized exoplanets, the system offers a unique opportunity to study terrestrial exoplanets orbiting an ultra-cool dwarf ($T_\textnormal{eff}=2550$ K, $R_* = 0.12R_\odot$, $M_*=0.09 M_\odot$) with a wide range of equilibrium temperatures, from Venus-like (TRAPPIST-1~b, c) to those of the Earth, Mars and beyond (TRAPPIST-1~d, e, f, g, h). Given the large amount of \textit{JWST} time invested in exploring this exoplanetary system, and the wide anticipation of its study with \textit{JWST} by the exoplanet community \citep[see, e.g.,][]{lj:2019, fauchez:2019, pid:2020, lin:2021, lin-kalt:2022, kriss-fort:2022, rotman:2023, meadows:2023}, we now turn our discussion to highlights on the exploration of the TRAPPIST-1 exoplanet system by the observatory. In particular, we focus this presentation on the constraints on the exoplanets TRAPPIST-1~b and TRAPPIST-1~c which, at the date of writing, are the only ones for which results have been published in the refereed literature (but we note \textit{JWST} observations exist for all the other exoplanets in the system). This, in turn, will allow us to introduce some of the main techniques used to characterize rocky exoplanet atmospheres with \textit{JWST}, as well as the possible challenges in this endeavor.

\subsubsection{Constraints on possible atmospheres on TRAPPIST-1~b}

The very first set of atmospheric constraints on the TRAPPIST-1 exoplanets with \textit{JWST} were presented in the work of \cite{greene:2023}, who studied 5 secondary eclipses of \hbindex{TRAPPIST-1~b} ($R_p = 1.1R_\oplus$, $M_p = 1.4M_\oplus$, $T_{\textnormal{eq}}=400$ K) obtained with MIRI/imaging at 15 $\mu$m. The objective: detecting \hbindex{thermal emission} from the exoplanet's dayside in order to constrain the prescence or absence of an atmosphere on this relatively warm world. A large eclipse depth would imply a very hot dayside, suggesting the exoplanet is re-irradiating back most of the energy the star is shining on it; strong evidence for the exoplanet having little to no atmosphere --- i.e., a \hbindex{bare rock} scenario. A small eclipse depth, on the other hand, would suggest a somewhat cooler dayside measured at this wavelength and thus suggest either: (a) a more efficient redistribution of energy throughout the planet, perhaps due to an atmosphere, (b) absorption by CO$_2$ in the atmosphere of the exoplanet, which has one of the largest absorption features precisely at 15 $\mu$m, (c) a ``shiny" surface or cloud scenario, on which high Bond albedo $A_B$ (i.e., reflective) clouds ($A_B \gtrsim 0.3$) or a surface ($A_B \lesssim 0.2$) prevents the planet from absorbing incoming energy, making it cooler, or (d) a mixture of a-c \citep[see, e.g.,][]{koll:2019, mansfield:2019, hammond:2025}. In their study, \cite{greene:2023} indeed detect thermal emission from TRAPPIST-1~b ---a first in this exoplanet system--- and showed the large eclipse scenario was indeed the case for these first TRAPPIST-1~b observations (see Figure \ref{fig:10}), suggesting a thick atmosphere on TRAPPIST-1~b was unlikely \citep[see also][for a detailed theoretical modelling of this data]{ih:2023}. This possibility was, indeed, consistent with transmission spectroscopy measurements using NIRISS/SOSS by \cite{lim:2023}.

The work of \cite{ducrot:2024} presents an additional interesting piece to the puzzle of TRAPPIST-1~b. This work presents 5 additional eclipses to those introduced in \cite{greene:2023} but at 12.8 $\mu$m. While together they are consistent with the interpretation of a possible bare rock or thin atmosphere on TRAPPIST-1~b by \cite{greene:2023} and, indeed, ultramafic rock surfaces fit the data well (see Figure \ref{fig:10}), this work also introduces another interesting possibility: hazes in a thick, CO$_2$-dominated atmosphere in TRAPPIST-1~b could \textit{in principle} create a temperature inversion and result in CO$_2$ in \textit{emission}, causing a large eclipse depth at 15 $\mu$m due to this feature, and a smaller eclipse depth at 12.8 $\mu$m as observed (Figure \ref{fig:10}, top panels, orange model). While the work of \cite{ducrot:2024} notes this scenario might be photo and thermochemically unlikely, it does showcase areas of further study motivated by \textit{JWST} observations, which is exactly what the observatory was set to open up.

\subsubsection{Constraints on possible atmospheres on TRAPPIST-1~c}

Using the same technique and wavelength range as the one showcased in \cite{greene:2023} for TRAPPIST-1~b, the work of \cite{zieba:2023} presents four 15 $\mu$m MIRI secondary eclipses with which thermal emission is detected on the warmer sibling exoplanet \hbindex{TRAPPIST-1~c} ($R_p=1.1R_\oplus$, $M_p=1.3M_\oplus$, $T_\textnormal{eq}=340$ K). These precise observations allowed to disfavor a Venus-like atmosphere at the 2.5$\sigma$-level (see Figure \ref{fig:10}, bottom panels). However, an atmosphere cannot be ruled out altogether by this data alone. Various types of 0.1-10 bar atmospheres dominated by CO$_2$, H$_2$O or O$_2$ are still consistent with the data and atmospheric escape models \citep{lincowski:2023, teixeira:2024}. In line with this argument, two transmission spectra obtained by NIRISS/SOSS and presented in the work of \cite{radica:2025} showcase how the transmission spectrum in the 0.7-3$\mu$m range is consistent with various types of atmospheres (also showcased in Figure \ref{fig:10}).  

\subsubsection{The challenge of the Transit Light Source effect in TRAPPIST-1 and beyond}

For rocky exoplanet science, transmission spectroscopy at first glance seems to be, in general, the best strategy for exoplanet atmospheric exploration. Just as it was shown in previous sections, the technique in principle allows to detect a wide range of atmospheric features for gas giants and sub-Neptunes alike. The transmission spectroscopy studies on TRAPPIST-1~b and TRAPPIST-1~c of \cite{lim:2023} and \cite{radica:2025} show, however, that the transmission spectra is severely distorted by stellar heterogeneities --- i.e., it is evident the \hbindex{TLS} effect is real, and impacting \textit{JWST} observations in particular of exoplanets orbiting \hbindex{M-dwarfs}, as predicted pre-launch \citep[see, e.g.][]{rackham:2017, iyer:2020, sag21}. 

While the TLS effect can be corrected at a certain level, as discussed in \cite{lim:2023} and \cite{radica:2025} there is no good handle on the \textit{accuracy} of those corrections for M-dwarfs, as ground-truth for what exactly causes heterogeneities in these cool stars is, to date, unclear. Even if we knew the underlying physical processes, our stellar modelling capabilities have limitations that cap our understanding of the TLS effect at a level important for \textit{JWST} transmission spectroscopy \citep{iyer:2023,rk:2024}. While stellar contamination is observed for Hot Jupiter exoplanet atmospheric \textit{JWST} studies orbiting from active G-dwarfs \citep[see, e.g.,][]{ft:2024} to M-dwarfs \citep[see, e.g.,][]{canas:2025}, these mostly impact as a confounding/bias factor on the abundance estimation of different elements, as the atmospheric features in those cases are large (several hundreds to thousands of ppm). For rocky exoplanets this is problematic, however, because features have amplitudes of a few hundreds-to-tens of ppm, and the objective of most observations is to actually \textit{detect} an atmosphere. This has triggered many studies to understand how to solve this in the context of \textit{JWST} observations, from understanding the spectra of spots themselves via, e.g, 3D magnetohydrodynamics models \citep[see, e.g.,][]{smitha:2025} or empirically \citep[e.g.,][]{berardo:2024}, to using transits of multiple planets to correct for this contamination directly from the data, like the \textit{JWST} NIRSpec/PRISM contemporaneous observations of TRAPPIST-1~b and c of \cite{rathcke:2025}. 

To date, the TLS problem continues to be a challenge for \textit{JWST} small exoplanet studies in particular around exoplanets orbiting M-dwarfs via transmission spectroscopy. This has put forward propositions in part of the exoplanet community to, when possible, begin their characterization via \textit{emission} photometry and/or spectroscopy \citep[see, e.g.,][]{framework:2024}, as done for TRAPPIST-1 in the studies of \cite{greene:2023}, \cite{zieba:2023} and \cite{ducrot:2024}, before performing detailed transmission spectroscopy studies. While emission studies on M-dwarfs might have their own challenges \citep[see, e.g.,][]{fauchez:2025}, solutions to those seem within reach. As we discuss in the next sub-section, emission measurements have actually turned out to be some of the most constraining when it comes to putting strong constraints on the prescence or abscence of an atmosphere in rocky exoplanets. We provide an overview of some results on other rocky exoplanets orbiting M-dwarfs next. 

\subsection{Can rocky exoplanets around M-dwarfs hold on to their atmospheres?}

One of the key highlights of \textit{JWST}'s exoplanet atmospheric exploration has been the search for atmospheres on rocky exoplanets orbiting M-dwarfs. Being the most abundant stars in the solar neighborhood, these worlds host the majority of rocky worlds in the Sun's vicinity, being thus key to understanding them \citep{henry:2006,dc:2013}. These systems, in turn, offer one of the best opportunities to characterize the atmospheres of small, cool ---some even potentially habitable--- worlds: the small sizes of their stars and close proximity to their host stars make them optimal to characterize via the transmission and emission spectroscopy techniques discussed throughout this Chapter. 

\begin{figure}
\includegraphics[scale=0.45]{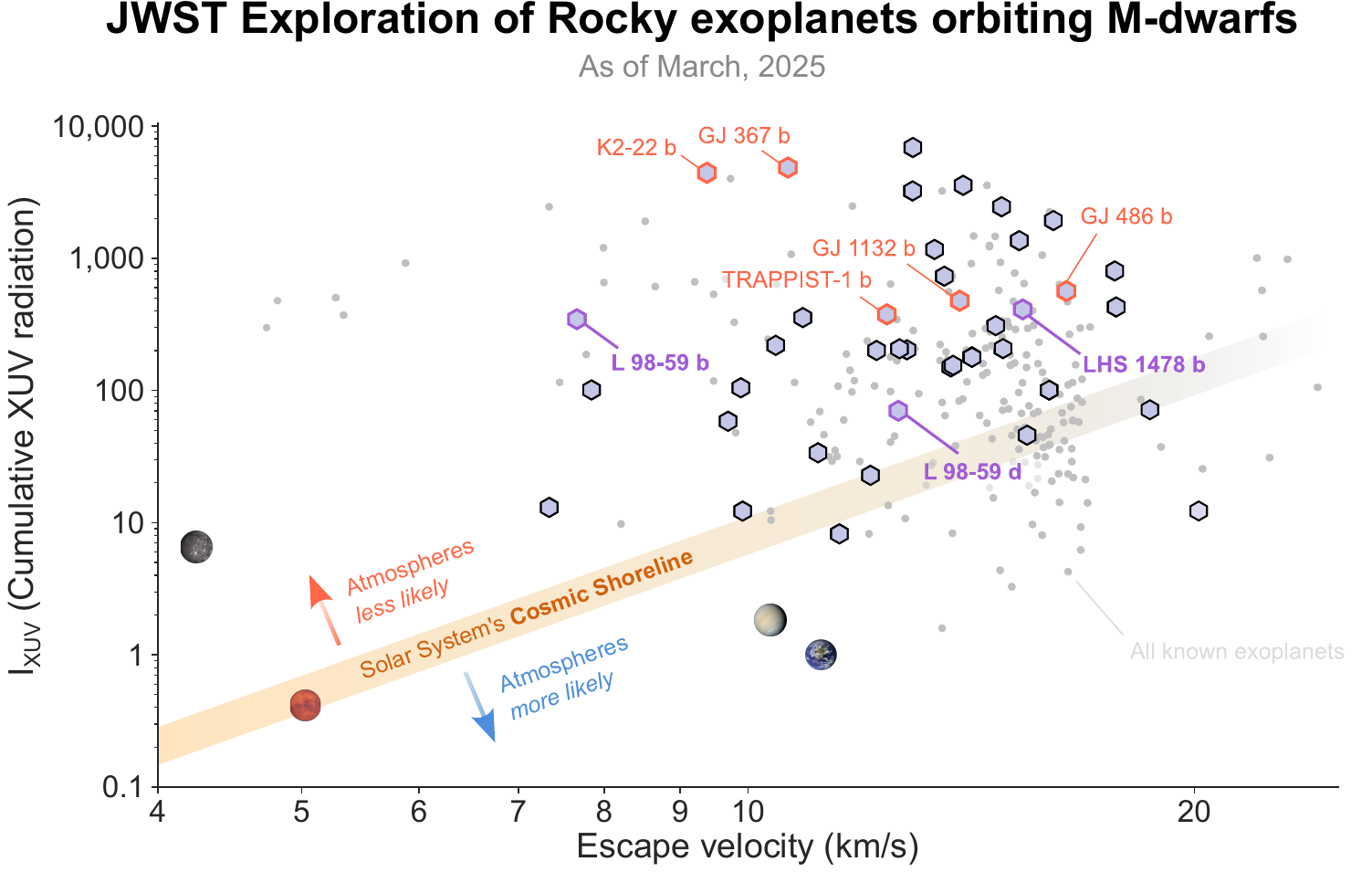}
\caption{\textbf{Cosmic shoreline diagram for rocky exoplanets orbiting M-dwarfs showcasing JWST observations and results.} Escape velocity in km/s versus estimated cumulative XUV radiation ---normalized to that of Earth's--- following \cite{cs:2017}, along with the Solar System's ``Cosmic Shoreline" (dividing line between Solar System objects with and without atmospheres) from that same work. Images of Solar System rocky planets (from left to right, Mercury, Mars, Venus, Earth) represent their position in the diagram. Rocky exoplanets (here, simply exoplanets $<2R_\oplus$) orbiting M-dwarfs (stars with $T_\textnormal{eff}<4000$ K) from the NASA Exoplanet Archive are shown as small, grey points. Objects being observed by \textit{JWST} are shown with pale purple hexagons. Exoplanets for which \textit{JWST} observations have shown to date atmospheres are unlikely are highlighted in red. Exoplanets for which \textit{JWST} observations may hint at the presence of an atmosphere, even if tentative at the time of writing, are highlighted in purple.}
\label{fig:11}       
\end{figure}

The question of whether these exoplanets can hold at all to their atmospheres in the harsh, high-activity M-dwarf environments they reside in has been identified as a key science theme for \textit{JWST}'s exoplanet atmospheric exploration \citep{redfield:2024}. To date, one of the most popular guiding principles for such searches is that of the ``\hbindex{Cosmic Shoreline}" --- an empirical line in the escape velocity versus the total stellar radiation intercepted by a planet (or the total cumulative XUV radiation intercepted by the planet) that, in the Solar System, divides planets with and without atmospheres \cite[see][and references therein]{cs:2017}. We show this diagram ---and the corresponding Cosmic Shoreline--- in Figure \ref{fig:11} along with all currently known exoplanets with radii smaller than $2R_\oplus$ orbiting M-dwarfs ($T_\textnormal{eff}<4000$ K). Exoplanets observed by \textit{JWST} are shown as hexagons. Whether rocky planets orbiting M-dwarfs also have Cosmic Shorelines similar (or even offset) to the one observed in the Solar System is one of the big questions \textit{JWST} observations are trying to answer.

\subsubsection{Constraints on rocky exoplanets with little or no atmospheres}

The very first set of highlights we introduce on this search for rocky exoplanet atmospheres around M-dwarfs are the exoplanets marked in red in Figure \ref{fig:11}, which represent objects for which \textit{JWST} has shown ample evidence \textit{against} a significant atmosphere. In addition to TRAPPIST-1~b, introduced in the previous sub-section, the \textit{JWST} MIRI emission and/or phase-curve observations of the exoplanets \hbindex{GJ 1132~b} \citep[$R_p=1.2R_\oplus$, $M_p=1.8M_\oplus$, $T_\textnormal{eq}=580$ K;][]{xue:2024}, \hbindex{GJ 486~b} \citep[$R_p=1.3R_\oplus$, $M_p=2.8M_\oplus$, $T_\textnormal{eq}=690$ K;][]{mansfield:2024} and \hbindex{GJ 367~b} \citep[$R_p=0.7R_\oplus$, $M_p=0.6M_\oplus$, $T_\textnormal{eq}=1,360$ K;][]{zhang:2024} are all inconsistent with thick ($\gtrsim 1-10$ bar) atmospheres, and instead favor thin atmospheres or bare rocks. These results are, in turn, consistent with the \textit{JWST} NIRSpec transmission spectroscopy studies for GJ 1132~b \citep{may:2023} and GJ 486~b \citep{moran:2023}, with the latter showing features in the transmission spectrum that the emission spectroscopy results of \cite{mansfield:2024} suggest are very likely produced by the TLS effect.

K2-22~b is a special case we would like to highlight on its own. This is a $<1 R_\oplus$ disintegrating exoplanet which has a rich history of observations showing variable transit events \citep[see, e.g.,][and references therein]{schlawin:2021}. While marked as having no atmosphere in Figure \ref{fig:11}, the \textit{JWST} MIRI observations in \cite{tusay:2025} do show some evidence for out-gassing material from this exoplanet. However, the pressures at which this gas would be is on the order of 10$^{-6}$ bar --- difficult to be classified as an ``atmosphere", but exciting for the prospects of figuring out processes occurring in a disintegrating exoplanet.

\subsubsection{Rocky exoplanets with hints of atmospheric features}

We have marked some exoplanets with purple (between ``red" and ``blue") in Figure \ref{fig:11}. These are objects for which there are \textit{hints} in current \textit{JWST} data for an atmosphere that only follow-up observations and analyses will be able to confirm. The first set of exoplanets we discuss here are the hints of atmospheres on \hbindex{L 98-59~b} \citep[$R_p=0.8R_\oplus$, $M_p=0.4M_\oplus$, $T_\textnormal{eq}=630$ K;][]{ba:2025} and its sibling planet \hbindex{L 98-59~d} \citep[$R_p=1.5R_\oplus$, $M_p=1.9M_\oplus$, $T_\textnormal{eq}=420$ K;][]{gressier:2024, banerjee:2024} --- both constrained via transmission spectroscopy with \textit{JWST} NIRSPec. The striking feature of those independent findings is the fact that for both planets, sulfur-rich atmospheres are suggested. In particular, the study of \cite{ba:2025} proposes an SO$_2$-dominated atmosphere best explains L 98-59~b's transmission spectrum, whereas the work of \cite{gressier:2024} and \cite{banerjee:2024} suggest that for L 98-59~d, SO$_2$ and/or H$_2$S could be the main atmospheric constituents. An important set of insights on these works is the fact that such atmospheres would need to have some sort of replenishment of sulfur to survive. For example, \cite{ba:2025} estimates that an SO$_2$-dominated atmosphere on L 98-59~b, given the high input XUV flux on the planet, should only last on the order of 10 Myr. Both works suggest that volcanic outgassing might be a reasonable explanation to the presence of those sulfur-rich atmospheres, if they are indeed real. 

The second exoplanet we highlight here is \hbindex{LHS 1478~b} ($R_p=1.2R_\oplus$, $M_p=2.3M_\oplus$, $T_\textnormal{eq}=590$ K), for which the work of \cite{august:2024} suggests a shallow secondary eclipse at 15 $\mu$m via \textit{JWST} MIRI photometric observations. This might, in turn, hint at the presence of atmospheric absorption by CO$_2$. While this constraint is based on a single datapoint at 15 $\mu$m, as the second observations analyzed in that work suffered from what appears to be relatively strong systematics that prevented for the secondary eclipse to be detected, it is relatively constraining of the possible atmospheric pressures, with low pressure, $<0.1$ bar atmospheres being disfavored by the data at the 2$\sigma$ level.

\subsubsection{Trends and future outlook on the search of atmospheres on rocky exoplanets around M-dwarfs}

We conclude this section by emphasizing that the highlights presented here represent only a small fraction of ongoing \textit{JWST} observations of rocky exoplanets orbiting M-dwarfs. This is, thus, just the beginning of the constraints and discoveries \textit{JWST} will provide for rocky exoplanet atmospheric science in current and future cycles. Notably, even the small sample of published results to date is enabling detailed predictions and statistical studies \citep[see, e.g.,][]{kt:2023,park-coy:2024, foley:2024} illustrating how these early findings are shaping our understanding of rocky exoplanets around M-dwarfs and driving the community forward, motivating detailed future studies. These efforts are laying the foundation for upcoming large-scale programs such as the Rocky Worlds Director's Discretionary Time (DDT) 500-hour \textit{JWST} / 250-orbit \textit{HST} program --- a key, specific recommendation from the Working Group on Strategic Exoplanet Initiatives with \textit{JWST} and \textit{HST}, identified through community input as the next crucial step in the search for evidence of atmospheres on rocky exoplanets orbiting M-dwarfs \citep{redfield:2024}. 

\section{Conclusions}

In this Chapter, we have summarized some of the highlights on the exploration of exoplanetary systems with \textit{JWST}. From gas giants to small, rocky exoplanets, the observatory has showcased to have a very versatile portfolio of instrumentation and state-of-the-art technology that is enabling cutting-edge scientific advancements in the field.

The exoplanetary science being performed on gas giant exoplanets by \textit{JWST} is vast. The observatory is allowing for the discovery of new exoplanets via high-contrast imaging techniques in stellar environments and mass-regimes that have been unexplored to date. It is also enabling detailed atmospheric characterization studies for directly imaged and transiting exoplanets, with which \textit{JWST} is revealing chemical inventories and physical processes with an unprecedented level of detail. Similarly, the technology onboard \textit{JWST} is also allowing for the detailed characterization of the sub-Neptune population ---the most numerous population in our galaxy for close-in orbits and--- one with no analogue in the Solar System. This exploration is unveiling how these exoplanets might have diverse internal structures, with some bridging to terrestrial worlds and others to gas giant exoplanets, breaking our Solar System intuition about their nature in the process. 

Finally, the study of rocky exoplanets by \textit{JWST} is enabling direct comparisons between rocky exoplanets elsewhere and those in our own Solar System, placing them in a broader cosmic context. The particular focus on studying rocky exoplanets orbiting M-dwarfs and whether they can retain their atmospheres serves as a crucial bridge to the imminent study of rocky exoplanets at Earth-like distances around Sun-like stars --- an objective that is fueling one of the main motivations for the next generation of space-based observatories, which is materializing in the \hbindex{Habitable Worlds Observatory} \citep{nat:2021}. \textit{JWST}, thus, provides a direct pathway to studying the prevalence of atmospheres like the ones observed in our Solar System accross different stellar environments. This key complement may not only inform future missions by revealing yet-to-be-understood atmospheric escape processes and/or chemical diversity in these rocky exoplanet atmospheres, but also deepen our understanding of our very own existence on a habitable planet like Earth.  

\section{Data and Figures availability}

All the editable versions of the Figures in this Chapter (along with data and \texttt{python} scripts to make them, when appropriate) can be found in \url{https://github.com/nespinoza/jwhighlights} \citep{jwrepo}. This includes a database with all the exoplanets being observed by \textit{JWST}, along with their physical properties, the ones for their stars, instrument/modes and observing program IDs. This data was manually digitized and cross-referenced to properties from the NASA Exoplanet Archive by NE. If you make use of this database or any of the scripts and data in the repository, please cite this book and chapter, along with the respective studies if using data from them. If you find any missing planets/programs, please open a Github issue.

\section{Cross-References}
\begin{itemize}
\item{Observing Exoplanets with the James Webb Space Telescope}
\item{Exoplanet Atmosphere Measurements from Transmission Spectroscopy and Other Planet Star Combined Light Observations}
\item{Characterization of Exoplanets: Secondary Eclipses}
\item{Exoplanet Phase Curves: Observations and Theory}
\item{Detecting and Characterizing Exomoons and Exorings}
\item{TRAPPIST-1 and its compact system of temperate rocky planets}
\end{itemize}

\begin{acknowledgement}
NE and MP would like to thank Jenny Novacescu and the Library team at STScI for insights and help looking for ways to \textit{not} miss references and studies for this Chapter. They would also like to thank the tens of thousands of people that made ---and keep making--- \textit{JWST} a reality through cooperation between NASA, STScI, CSA, ESA and other institutional partners, showcasing how humankind, working together, can accomplish the seemingly impossible. NE would like to thank the Transiting Exoplanet Group at STScI for general feedback on figures made for this Chapter. NE would also like to thank conversations on style and presentation of material with H. Diamond-Lowe as well as with M. P. Mart\'inez. NE would like to thank several members of the STScI/JHU community on very enjoyable conversations on ``what is an exoplanet"; in particular, with A. Carter, W. Balmer and J. Espinoza. NE would like to thank A. Carter and H. Diamond-Lowe for early feedback on sections of this Chapter. NE would like to acknowledge the Inkscape developers that allowed many of the graphics in this Chapter to happen. This research has made use of the NASA Exoplanet Archive, which is operated by the California Institute of Technology, under contract with the National Aeronautics and Space Administration under the Exoplanet Exploration Program.
\end{acknowledgement}
\pagebreak

\end{document}